\documentclass[aps,prb,reprint,onecolumn,11pt,groupedaddress]{revtex4-2}
\usepackage[colorlinks=true,linkcolor=blue,citecolor=blue,urlcolor=blue]{hyperref}
\usepackage{amsmath}
\usepackage{amssymb} 
\usepackage[pdftex]{graphicx}
\usepackage{braket} 

\newcounter{firstbib}

\begin{document}


\title{Quantum non-demolition measurement of an electron spin qubit through its low-energy many-body spin environment}



\author{Harry E. Dyte}
\affiliation{Department of Physics and Astronomy, University of
Sheffield, Sheffield S3 7RH, United Kingdom}
\author{George Gillard}
\affiliation{Department of Physics and Astronomy, University of
Sheffield, Sheffield S3 7RH, United Kingdom}
\author{Santanu Manna}
\affiliation{Institute of Semiconductor and Solid State Physics,
Johannes Kepler University Linz, Altenberger Str. 69, 4040 Linz,
Austria}
\author{Saimon F. Covre da Silva}
\affiliation{Institute of Semiconductor and Solid State Physics,
Johannes Kepler University Linz, Altenberger Str. 69, 4040 Linz,
Austria}
\author{Armando Rastelli}
\affiliation{Institute of Semiconductor and Solid State Physics,
Johannes Kepler University Linz, Altenberger Str. 69, 4040 Linz,
Austria}
\author{Evgeny A. Chekhovich}
\email[]{e.chekhovich@sheffield.ac.uk} \affiliation{Department of
Physics and Astronomy, University of Sheffield, Sheffield S3 7RH,
United Kingdom}

\date{\today}

\begin{abstract}
The measurement problem dates back to the dawn of quantum mechanics. Here, we measure a quantum dot electron spin qubit through off-resonant coupling with thousands of redundant nuclear spin ancillae. We show that the link from quantum to classical can be made without any ``wavefunction collapse'', in agreement with the Quantum Darwinism concept. Large ancilla redundancy allows for single-shot readout with high fidelity $\approx99.85\%$. Repeated measurements enable heralded initialization of the qubit and probing of the equilibrium electron spin dynamics. Quantum jumps are observed and attributed to burst-like fluctuations in a thermally populated phonon bath.
\end{abstract}


\maketitle

\begin{figure}
\includegraphics[width=0.6\linewidth]{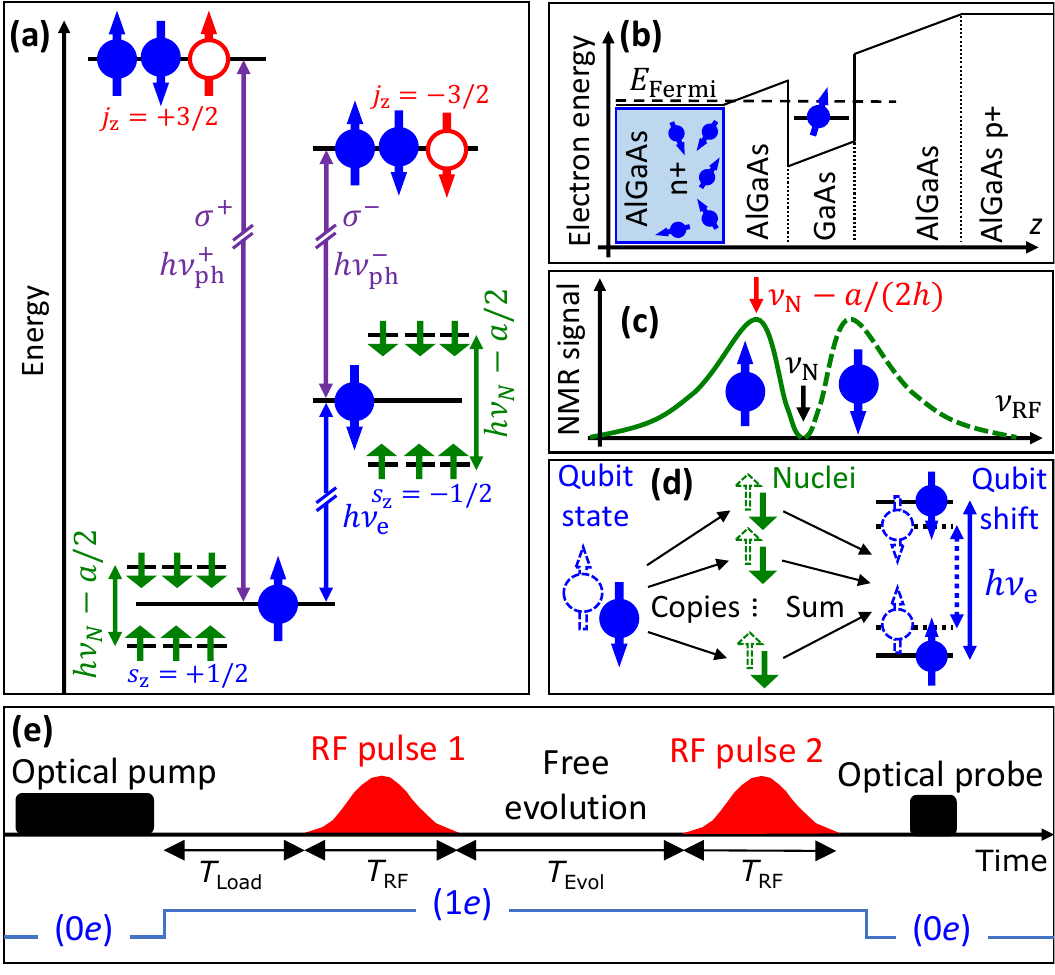}
\caption{(a) Energy level diagram of the nuclear spins (short arrows), the electron (solid circle and arrow) and the optically excited trion, containing two electrons and one hole (open circle and arrow). (b) Conduction band energy diagram of the semiconductor structure, showing GaAs quantum dot, AlGaAs barriers and the doped AlGaAs layers. (c) The solid (dashed) line shows schematically the nuclear magnetic resonance (NMR) spectrum in presence of a spin-up (spin-down) electron. Vertical arrows show the bare nuclear frequency $\nu_{\rm{N}}$ and the frequency $\nu_{\rm{N}}-a/(2h)$ of the detuned radiofrequency (RF) pulse. (d) The electron spin qubit projection is first copied into multiple nuclear spin ancillae by the RF pulse. The total nuclear polarization is then measured from the hyperfine shift of the electron spin qubit. (e) Timing diagram showing optical pump and probe, RF pulses and the switching of the QD between the neutral (0$e$) and electron-charged (1$e$) sates.} \label{Fig:Intro}
\end{figure}

\begin{figure}
\includegraphics[width=0.6\linewidth]{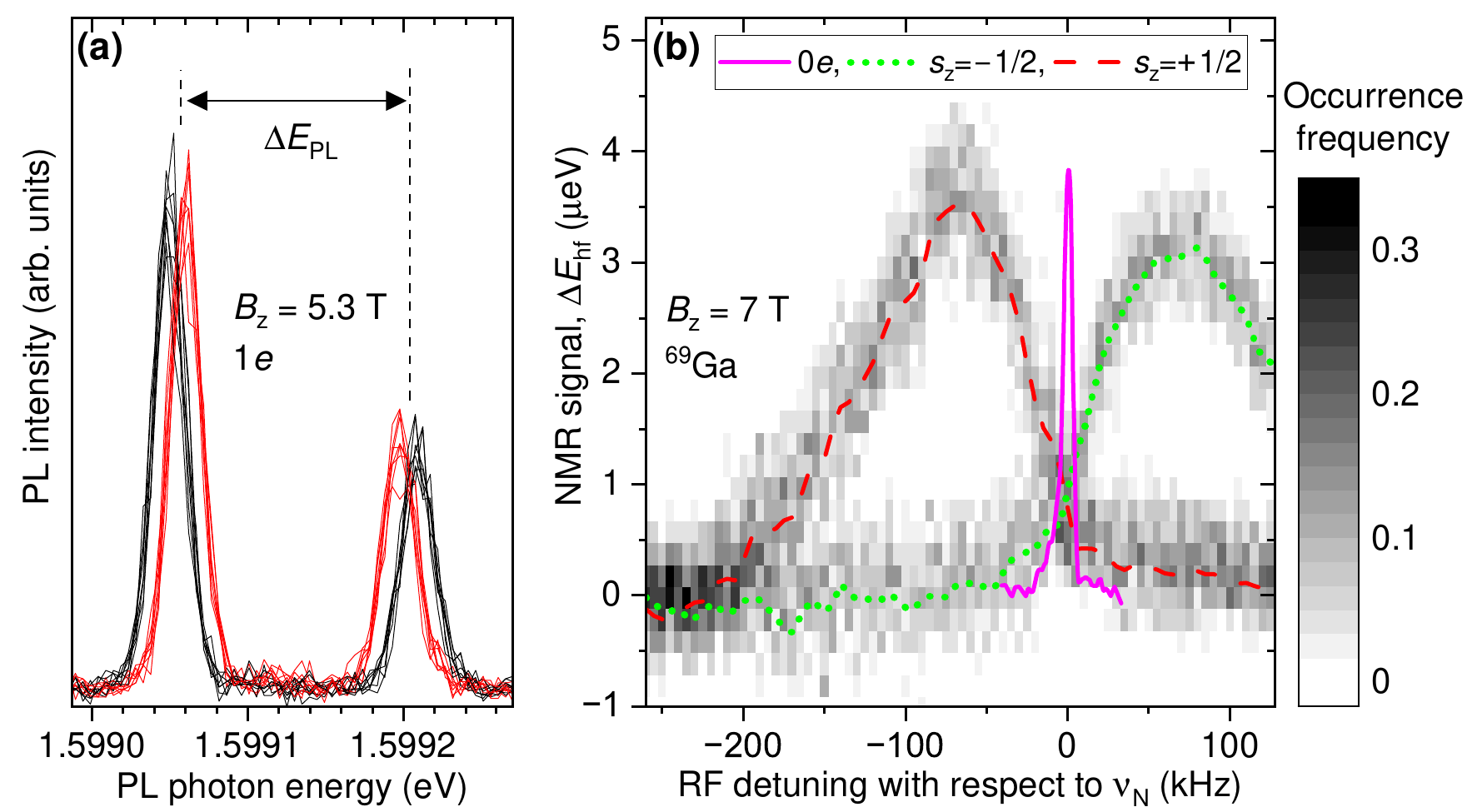}
\caption{(a) A random set of 16 probe PL spectra following the detuned RF $\pi$ pulse applied to a charged (1$e$) QD. The changes in the doublet splitting $\Delta E_{\rm{PL}}$ are due to the changes $\Delta E_{\rm{hf}}$ in the hyperfine shift $E_{\rm{hf}}$. The bimodality in $\Delta E_{\rm{PL}}$ (and $\Delta E_{\rm{hf}}$) corresponds to the two $s_{\rm{z}}$ states of the electron. (b) Histogram of the single-shot NMR signals $\Delta E_{\rm{hf}}$ measured with an RF $\pi$ pulse at variable detunings from the  $^{69}$Ga bare NMR frequency $\nu_{\rm{N}}$ ($\nu_{\rm{N}}\approx72.15$~MHz at $B_{\rm{z}}\approx7$~T). The dashed (dotted) line traces the branch of the NMR resonance corresponding to the $s_{\rm{z}}=+1/2$ ($-1/2$) electron spin state. The solid line shows the same single-QD resonance but measured in a neutral charge state (0$e$) via ``inverse'' NMR method \cite{Chekhovich2012}.} \label{Fig:SingleShot}
\end{figure}

\begin{figure}
\includegraphics[width=0.6\linewidth]{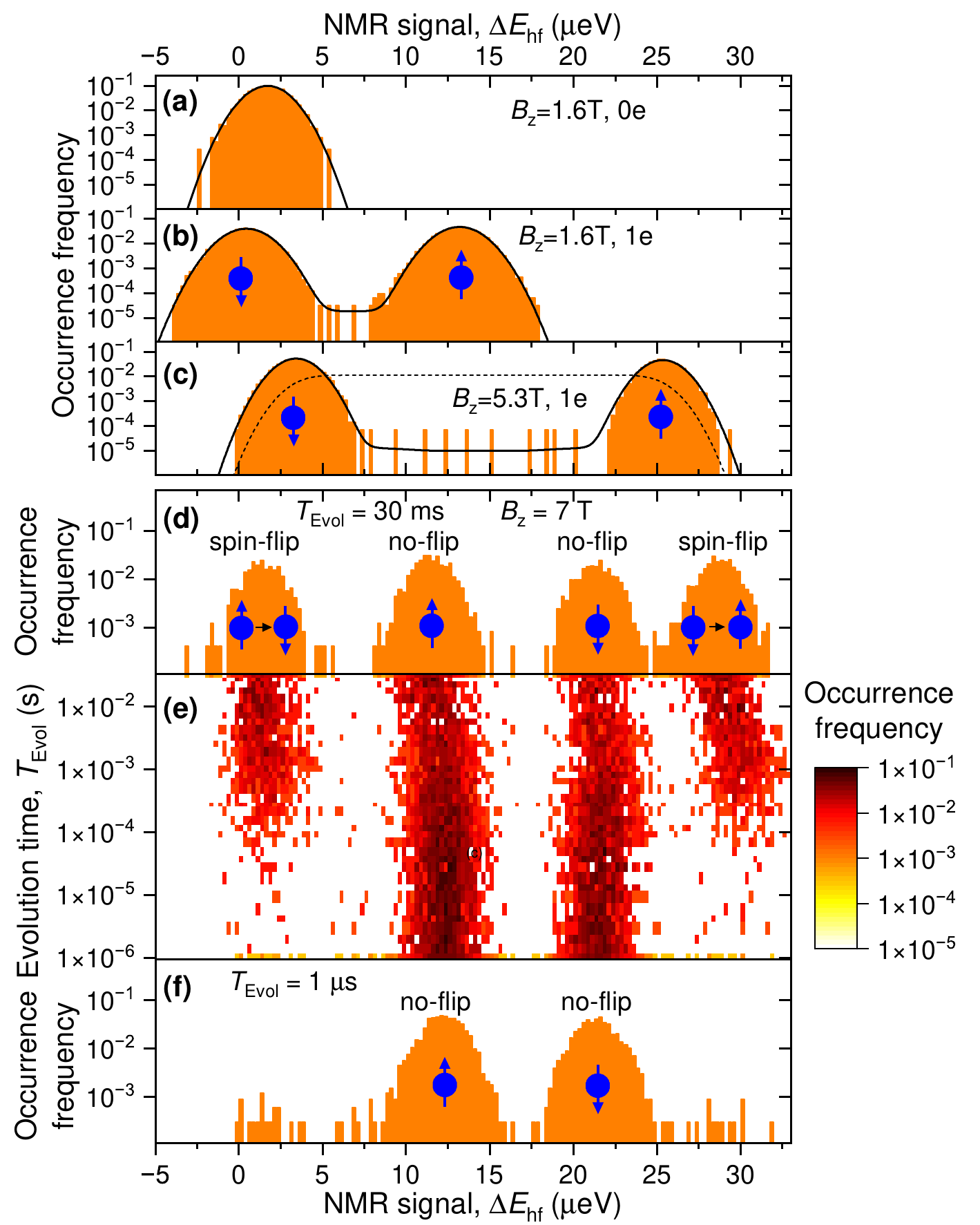}
\caption{(a-b) Histograms of the single-shot NMR signals $\Delta E_{\rm{hf}}$ measured at $B_{\rm{z}}=1.6$~T on the same individual QD in a neutral charge state (0$e$, a), and in a single-electron charged state (1$e$, b). The NMR signals are produced by a single detuned RF $\pi$ pulse. Solid lines show the best model fitting. (c) Same as (b) but on a different QD and at $B_{\rm{z}}=5.3$~T. The dashed line shows a model distribution for a randomly oriented electron spin. (d--e) Histograms of the single-shot NMR signals $\Delta E_{\rm{hf}}$ measured at $B_{\rm{z}}=7$~T with two RF $\pi$ pulses applied to $^{75}$As and $^{69}$Ga and delayed by $T_{\rm{Evol}}$. A full 2D histogram at variable $T_{\rm{Evol}}$ is shown in (e), while (d) and (f) show the cross sections at long and short $T_{\rm{Evol}}$, respectively.} \label{Fig:Hist}
\end{figure}

\newcommand{\FigBand}{Fig.~\ref{Fig:Intro}(b)}
\newcommand{\FigEDiag}{Fig.~\ref{Fig:Intro}(a)}
\newcommand{\FigQDarwin}{Fig.~\ref{Fig:Intro}(d)}
\newcommand{\FigTDiag}{Fig.~\ref{Fig:Intro}(e)}
\newcommand{\FigRFMmt}{Fig.~\ref{Fig:Intro}(c)}
\newcommand{\FigPL}{Fig.~\ref{Fig:SingleShot}(a)}
\newcommand{\FigNMReZ}{Fig.~\ref{Fig:SingleShot}(b)}
\newcommand{\FigHistZ}{Fig.~\ref{Fig:Hist}(a)}
\newcommand{\FigHistLowB}{Fig.~\ref{Fig:Hist}(b)}
\newcommand{\FigHistHighB}{Fig.~\ref{Fig:Hist}(c)}
\newcommand{\FigTEvolHighBShort}{Fig.~\ref{Fig:Hist}(d)}
\newcommand{\FigTEvolHighBLong}{Fig.~\ref{Fig:Hist}(f)}
\newcommand{\FigTEvolHighB}{Fig.~\ref{Fig:Hist}(e)}

High fidelity qubit readout is essential in quantum information processing. Usually, such readout starts with conversion of a fragile quantum state into a more robust form, detectable by a classical apparatus. Some readout techniques rely on high-energy excitations, making this conversion dissipative (irreversible). Examples include spin-to-charge conversion \cite{Elzerman2004, Hensen2020, Meunier2006, Veldhorst2014}, single photon detection \cite{Hadfield2009}, optical readout of spin in defects \cite{Jiang2009,Robledo2011,Raha2020, Kindem2020, Evans2018, Bhaskar2020} and quantum dots (QDs) \cite{Vamivakas2010, Delteil2014, Antoniadis2022}. An alternative is unitary (reversible) conversion. One example is the off-resonant (Ising) coupling between the main and ancilla electron spin qubits, which enables quantum non-demolition (QND) measurement \cite{Yoneda2020}. Other QND demonstrations include superconducting qubits \cite{Blais2004} and mechanical resonators \cite{Rossi2018}.

Here, we implement unitary conversion of a QD electron spin, but the ancilla is of a different nature, consisting of $\approx10^4-10^5$ low-energy nuclear spin qubits. The large redundancy of the ancilla results in a very high readout fidelity, which is what an observer perceives as a deterministic classical measurement. The only departure from an ideal quantum-to-classical conversion comes from random qubit jumps. However, unlike in previous studies \cite{Delteil2014,Raha2020}, the jumps are not caused by the measurement itself. Instead, the electron spin jumps are attributed to spontaneous bursts of electric fields, produced by the equilibrium vibrations of the crystal lattice (phonons). Our readout method is particularly robust and simple to implement, since the nuclei are essentially the same in all QDs, eliminating the need for QD-specific calibrations.

We study lattice-matched epitaxial GaAs QDs grown by in-situ etching and infilling of nanoholes in AlGaAs \cite{Heyn2009,Atkinson2012,Gurioli2019,Gillard2022,Zaporski2022}. The QD can be charged with a single electron from the $n$-type Fermi reservoir, by adjusting the bias in a $p$-$i$-$n$ diode structure [\FigBand]. A static magnetic field $B_{z}$ is applied along the growth axis \textit{z}. A typical QD consists of  $N\approx10^5$ atoms, whose nuclei are spin-3/2 particles. The sample is subject to uniaxial stress, which induces nuclear quadrupolar shifts. This way the two-level subspace with nuclear spin projections $I_{\rm{z}}={-3/2,-1/2}$ is isolated, allowing the nuclei to be treated as spin-1/2 particles. Individual QDs are addressed optically using focused laser excitation and photoluminescence (PL) spectroscopy. A copper coil is used to generate a radiofrequency (RF) magnetic field orthogonal to $B_{z}$. Further details can be found in Supplementary.

The quantum system of a QD charged with a single electron (1$e$) is described with reference to the level diagram in \FigEDiag. The hyperfine interaction Hamiltonian is $\mathcal{H}_{\rm{hf}} = \Sigma_{k}a_{k}\hat{\textbf{s}} \cdot \hat{\textbf{I}}_{k}$, where $a_{k}$ describes the coupling between the spin vector $\textbf{s}$ of the resident electron and the $k$-th nuclear spin vector $\textbf{I}_{k}$. This interaction has a twofold effect. Firstly, in addition to the bare Larmor frequency $\nu_{\rm{N}}$, each nucleus acquires a Knight \cite{Knight1949} frequency shift $s_{\rm{z}} a_{k}/(2h)$. Secondly, the electron states with $s_{\rm{z}}=\pm1/2$ acquire the (Overhauser) hyperfine shifts $\pm E_{\rm{hf}}/2$, arising from the net polarization of the nuclear spin ensemble [\FigQDarwin]. The average hyperfine shift is defined as $E_{\rm{hf}}=\Sigma_{k}a_{k}\langle\hat{I}_{\rm{z,k}} \rangle$, where $\langle...\rangle$ is the expectation value. The electron spin energy splitting $h\nu_{\rm{e}}$, is the sum of $E_{\rm{hf}}$ and the bare Zeeman splitting $h\nu_{{\rm{e}},0}=\mu_{\rm{B}} g_{\rm{e}} B_{\rm{z}}$, where $g_{\rm{e}}$ is the electron $g$-factor and $\mu_{\rm{B}}$ is the Bohr magneton. The optically excited trion, contains a spin-singlet pair of electrons and an unpaired valence band hole with momentum projection $j_{\rm{z}}=\pm3/2$. Due to the selection rules, there are two dipole-allowed circularly polarized ($\sigma^\pm$) optical transitions with photon energies $h\nu_{\rm{ph}}^{\pm}$. The optically-detected spectral splitting  $\Delta E_{\rm{PL}} = h(\nu_{\rm{ph}}^{+}-\nu_{\rm{ph}}^{-})$ yields the hyperfine shift $E_{\rm{hf}}$, up to a constant offset \cite{Urbaszek2013}. 

The traditional readout uses a cyclic optical transition [e.g. $\sigma^+$ in \FigEDiag] to convert the electron spin state into the presence or absence of scattered photons \cite{Vamivakas2010, Delteil2014, Antoniadis2022}. However, there is a finite probability for the measurement process to destroy the spin qubit if the recombination goes via one of the ``forbidden'' channels  [e.g. from $j_{\rm{z}}=+3/2$ to $s_{\rm{z}}=-1/2$ in \FigEDiag]. Here, we take a different approach, using the long coherence of the nuclear spins \cite{Gillard2022} and the large disparity of the energy scales $\nu_{\rm{ph}}^{\pm}\gg\nu_{\rm{e}}\gg\nu_{\rm{N}}$ to turn the nuclei into a non-invasive measurement apparatus. 

\FigTDiag\; shows the timing diagram of the measurement cycle. It starts with a long (few seconds) circularly-polarized optical pumping of an empty (0$e$) QD, which polarizes the nuclear spins up to $\approx80\%$ \cite{Chekhovich2017, MillingtonHotze2023}. Next, an electron is loaded from the Fermi reservoir (1$e$) and is allowed to equilibrate for a time $T_{\rm{Load}}$. Nuclear magnetic resonance (NMR) is performed by applying an RF pulse with a total duration $T_{\rm{RF}}$, calibrated to induce a $\pi$ rotation of the nuclear spins. In some experiments, a second RF pulse is applied, following a free evolution time $T_{\rm{Evol}}$. The final step is the illumination of the QD with a short (tens of milliseconds) optical probe in order to collect the PL spectrum and derive $E_{\rm{hf}}$. Importantly, all measurements are done in one cycle (i.e. single-shot), thus avoiding any averaging.

The readout of the electron spin qubit is explained in \FigRFMmt. An electron in state $s_{\rm{z}}=-1/2$ ($s_{\rm{z}}=+1/2$) Knight-shifts the QD NMR spectrum to the higher (lower) frequency side of $\nu_{\rm{N}}$. A single RF pulse is applied at a radiofrequency $\nu_{\rm{N}}-a/(2h)$, where $a$ is a weighted average of $a_{k}$ in a QD. For the electron in the $s_{\rm{z}}=+1/2$ ($-1/2$) state, the RF pulse is in (out of) resonance, so the QD nuclei are flipped (remain in the initial state) \cite{Jiang2009}. Statistics of the single-shot PL probe spectra [\FigPL] show a clear bimodality in the spectral splitting (red and black traces), arising from bimodal distribution of the RF-induced hyperfine shifts $\Delta E_{\rm{hf}}$. A systematic dependence of $\Delta E_{\rm{hf}}$ on the RF detuning from $\nu_{\rm{N}}$ is shown in \FigNMReZ, where the two branches corresponding to $s_{\rm{z}}=+1/2$ and $-1/2$ are traced by the dashed and dotted lines, respectively. The broadening of these traces arises from the inhomogeneous distribution of $a_{k}$, whereas the empty-QD (0$e$) NMR spectrum is much narrower (solid line). The optimal resolution of the two electron spin states (the maximum difference in $\Delta E_{\rm{hf}}$) is observed when the RF detuning matches the typical Knight shift $a/(2h)\approx70$~kHz.

Using the optimal detuning, we collect detailed statistics of the single-shot $\Delta E_{\rm{hf}}$. In an empty QD [0$e$, \FigHistZ\;] the distribution of $\Delta E_{\rm{hf}}$ is a single mode, broadened by the noise in probe PL spectra. The mode is centred at a small value $\Delta E_{\rm{hf}}\approx1.7~\mu$eV, indicating partial rotation of the nuclei by the detuned RF pulse. The same measurement in a charged QD [1$e$, \FigHistLowB] shows a bimodal distribution. One mode is centered at $\Delta E_{\rm{hf}}\approx0.4~\mu$eV and corresponds to the $s_{\rm{z}}=-1/2$ electron state, which Knight-shifts the nuclei out of resonance with the RF pulse. The mode at $\Delta E_{\rm{hf}}\approx13.2~\mu$eV corresponds to the $s_{\rm{z}}=+1/2$ state, which brings the nuclei into resonance with the RF pulse.

These results match the Quantum Darwinism perspective \cite{Zurek2018}, which recognizes that a direct measurement of a qubit is rarely possible. Instead, the observer uses the environment to acquire information about the qubit states indirectly. The observer then relies on a large number of redundant copies in order to arrive at the classical (deterministic) notion of objective reality \cite{Zurek2009}. In our experiments, the nuclear spin ensemble is such an environment. The RF pulse is essentially an electron-controlled CNOT gate acting simultaneously on multiple nuclear qubits \citep{Yang2010} to copy the electron state $s_{\rm{z}}$ into thousands of nuclear states $I_{{\rm{z}},k}$. At the final step, the gate bias ejects the electron from the QD, thus disconnecting the measurement apparatus (the nuclei) from the qubit. Illumination by the probe laser gradually destroys the individual nuclear spin copies, but their arithmetic sum [see \FigQDarwin] is robust enough to collect thousands of PL photons and measure the hyperfine shift $E_{\rm{hf}}$. This summation of redundant copies into essentially a classical variable (nuclear magnetization) is what enables a single-shot measurement of a quantum variable $\hat{s}_{\rm{z}}$ by a classical instrument (optical spectrometer with a photo-detector).

Returning to the single-shot NMR histograms, we note a small number of events where the NMR signal deviates from either of the modes [$8~\mu{\rm{eV}}\lesssim\Delta E_{\rm{hf}}\lesssim21~\mu{\rm{eV}}$ in \FigHistHighB]. We ascribe such intermediate readouts to electron spin flips during the RF pulse, resulting in partial rotation of the nuclear spins. We model this process by assuming a probability $p_{\rm{Flip}}$ for the electron spin to be flipped during $T_{\rm{RF}}$, leaving a probability $1-p_{\rm{Flip}}$ for the electron spin to maintain its $s_{\rm{z}}$. The optical readout noise is also included in the full model (see Supplementary Information). The best-fit results are shown by the solid lines in Figs.~\ref{Fig:Hist}(b) and \ref{Fig:Hist}(c). Using the fitted mode positions $\Delta E_{\rm{hf}}^-$ and $\Delta E_{\rm{hf}}^+$, we set the detection threshold in the middle $(\Delta E_{\rm{hf}}^-+\Delta E_{\rm{hf}}^+)/2$ and calculate the probability that the detected $\Delta E_{\rm{hf}}$ is below (above) the threshold when the true electron state is $s_{\rm{z}}=-1/2(+1/2)$. This probability is the qubit readout fidelity, found to be $F\approx0.9985$, matching or exceeding the state of the art in a range of qubit systems \cite{Hensen2020, Zhang2021, Raha2020, Evans2018, Bhaskar2020}. Since the two histogram modes are well resolved, the loss of fidelity is dominated by the random electron spin flips, leading to $F \approx 1-p_{\rm{Flip}}/2 \approx 1-T_{\rm{RF}}/(4T_{1,\rm{e}})$, where $T_{1,\rm{e}}$ is the electron spin lifetime. Resolution of the $s_{\rm{z}}=\pm1/2$ Knight-shifted NMR spectra with a short (spectrally broad) RF pulse, imposes the lower limit $T_{\rm{RF}}\gtrsim h/a$, where $a\propto N^{-1}$. Our experiments with $T_{\rm{RF}}\approx10-20~\mu$s are already close to the lower limit imposed by $N\approx 10^5$ in the studied QDs. On the other hand, nuclear spin relaxation \cite{MillingtonHotze2022} and decoherence \cite{Gillard2022} times are much longer than $h/a$ and therefore do not limit $F$. The other limitation comes from $T_{1,\rm{e}}$, which ranges from milliseconds to tens of milliseconds for temperature $T\approx4.2$~K and our typical electron spin splitting $h\nu_{\rm{e}}\approx50~\mu$eV. Further increase in $T_{1,\rm{e}}$ (and hence increase in $F$) can be achieved by lowering the temperature towards $k_{\rm{B}}T\approx h\nu_{\rm{e}}$, and by lowering $h\nu_{\rm{e}}$ through reduced magnetic field and nuclear spin polarization.

Immediate repeatability is a key requirement for any quantum measurement \cite{Ralph2006,Zurek2018}, which we verify in an experiment with two RF pulses [\FigTDiag]. The first pulse applied to $^{75}$As nuclei records the initial state, while the second pulse on $^{69}$Ga stores the state after the interpulse delay $T_{\rm{Evol}}$. The optically-measured $\Delta E_{\rm{hf}}$ is the total NMR signal produced by the two pulses. \FigTEvolHighB\; shows a two-dimensional histogram of $\Delta E_{\rm{hf}}$ measured at different $T_{\rm{Evol}}$. A cross-section at short $T_{\rm{Evol}}\approx1~\mu$s [\FigTEvolHighBShort] reveals the same bimodal distribution as in Figs.~\ref{Fig:Hist}(b) and \ref{Fig:Hist}(c), with only two ``no-flip'' modes corresponding to $s_{\rm{z}}=\pm1/2$. The two additional ``spin-flip'' modes, corresponding to $s_{\rm{z}}$ inversion during $T_{\rm{Evol}}$, emerge only at long $T_{\rm{Evol}}$ [$\approx30$~ms in \FigTEvolHighBLong]. Analysis of the entire $T_{\rm{Evol}}$ dependence reveals the spin lifetime $T_{1,\rm{e}}\approx0.58$~ms at $B_{\rm{z}}=7$~T, measured in equilibrium without any active initialization of the electron spin. Instead, a heralded initialization is performed by the first RF pulse, which stores the initial electron state in the $^{75}$As polarization, to be retrieved by the optical probe afterwards. The repeatability in the two-pulse experiments further highlights the unitarity of the measurement process -- although the final ejection of the electron can be seen as a qubit ``collapse'', it only occurs after $s_{\rm{z}}$ has been measured and recorded in the nuclei. The unitarity in our system is made possible by the low energy of the nuclei and the excellent fidelity of the RF coherent control, arising from a precise description of the microscopic electron-nuclear interactions. We argue that the non-unitary ``wavefunction collapse'' can be a mere simplification, invoked when the microscopic picture is missing (for example if the measurement involves coupling of the qubit to a high-energy environment).

The readout time $T_{\rm{RF}}=20~\mu$s is short enough to follow the electron spin evolution on the timescale of $T_{1,\rm{e}}$. However, \FigTEvolHighB\; shows that the electron spin is nearly always detected in either of the eigenstates $s_{\rm{z}}=\pm1/2$, with very rare intermediate NMR readouts $\Delta E_{\rm{hf}}$. This can be explained only if evolution of the electron spin is a random telegraph process, where the electron is in one of the eigenstates $s_{\rm{z}}=\pm1/2$ most of the time, occasionally experiencing quantum jumps (that are much faster than $T_{\rm{RF}}$).

We identify the origin of the jumps by combining our experiments with the first-principle numerical modelling, where the Schrödinger equation is propagated from the initial wavefunction state $\psi_{\rm{Init}}$ into the final state $\psi_{\rm{Fin}}$ (see details in Supplementary). We simulate the measurement process by initializing the nuclei (up to $N=12$) into a polarized state and initializing the electron spin in an arbitrary superposition $\psi_{\rm{Init}}=\alpha \vert+1/2\rangle + \beta \vert-1/2\rangle$ with the $z$-projection expectation value $s_{\rm{z,Init}}=(\vert\alpha\vert^2-\vert\beta\vert^2)/2$. Following the RF measurement pulse, we find that (i) the final polarization of each nucleus equals the initial electron polarization $I_{{\rm{z}},k,{\rm{Fin}}}\approx s_{\rm{z,Init}}$ and (ii) the electron polarization is nearly unchanged $s_{\rm{z,Fin}}\approx s_{\rm{z,Init}}$. Such non-demolition copying of the quantum variable $\hat{s}_{\rm{z}}$ comes at the expense of completely erasing the conjugate variable \cite{Ralph2006}, which manifests in $s_{\rm{x,Fin}}\approx s_{\rm{y,Fin}}\approx 0$ regardless of $s_{\rm{x,Init}}$. This result can be understood qualitatively through the large difference in the nuclear and electron precession frequencies $\nu_{\rm{N}}\ll \nu_{\rm{e}}$, meaning that the nuclei sense only the average electron polarization $\langle s_{\rm{z}} \rangle$. Moreover, the disparity in the energy scales $N\nu_{\rm{N}}<\nu_{\rm{e}}$ means that the electron follows adiabatically the evolution of the nuclear spin polarization \cite{Merkulov2002}. In other words, the nuclei rotated by the RF do not have enough energy to flip the electron spin, ensuring the QND nature of the measurement.

The linear response of the measurement apparatus $I_{{\rm{z}},k,{\rm{Fin}}}\approx s_{\rm{z,Init}}$, revealed by numerical modelling, provides the following insight into the origin of quantum jumps. If all electron spin superpositions had equal probabilities, the single-shot NMR signals would have had a uniform distribution, calculated and shown by the dashed line in \FigHistHighB. And yet the measurements yield a sharp bimodal distribution, revealing the energy eigenstates $s_{\rm{z}}=\pm1/2$ as a preferential basis. Quantum mechanics does not prescribe any preferential eigenbasis towards which the superpositions should decohere. Such a preferential basis can arise from the interaction of the qubit with the environment, known as einselection \cite{Schlosshauer2005,Zurek2018}. The nuclear spin environment has been ruled out above -- its energy is too small to ``project'' the high-energy electron spin qubit into the $s_{\rm{z}}=\pm1/2$ eigenstates. By contrast, the lattice vibrations (phonons) can act as a high-energy environment, leading to einselection and quantum jumps.
 
The inverse dependence of $T_{\rm{1,e}}$ on $B_{\rm{z}}$ (see Supplementary) confirms the dominant role of the phonons \cite{Khaetskii2001,Gillard2021}. The effective spin-phonon coupling is $\propto(\hat{s}_{\rm{x}}\mathcal{E}_{\rm{y}}-\hat{s}_{\rm{y}}\mathcal{E}_{\rm{x}})$, where $\mathcal{E}_{\rm{x,y}}$ are the Cartesian components of the phonon-induced piezo-strain electric field \cite{Khaetskii2001}. This spin-resonance form suggests that electron spin quantum jumps and einselection are driven by quasi-resonant electric fields, occurring in the form of short ($\ll 10~\mu$s) random bursts, separated by long (milliseconds) random intervals. Notably, these jumps are a spontaneous equilibrium process, as opposed to previous studies \cite{Delteil2014,Raha2020}, where the observation process (continuous optical excitation) could itself induce the qubit jumps. Spontaneous collapses and burst-like revivals have been investigated in Bosonic system, such as photons \cite{Eberly1980} and phonons \cite{Hizhnyakov1996,Misochko2004}, and are typically associated with high mode population numbers $\bar{n}\gtrsim 100$. The appearance of spontaneous revivals at the much lower average phonon numbers $\bar{n}\approx6.8$ (for $T=4.2$~K and $h\nu_{\rm{e}}\approx50~\mu$eV used here) is somewhat unexpected, calling for further experiments at variable $T$ and $h\nu_{\rm{e}}$. Higher-order correlations of the electron spin quantum jumps can be studied using three or more readout pulses. Sensitive detection of the low-energy phonons is itself an interesting application in the context of particle detection \cite{Young1992} and dark matter search \cite{Alkhatib2021}.

Finally, recent studies on the same GaAs QDs \cite{Zaporski2022} have revealed electron spin coherence times as long as $\approx 100~\mu$s, significantly exceeding our measurement time $T_{\rm{RF}}\approx 10~\mu$s. Thus, this QND readout method should allow for single-shot probing of the electron spin coherence without the need for dynamical decoupling, required in time-averaged measurements. Conversely, a detuned RF pulse can be used to generate and study the Greenberger–Horne–Zeilinger (Schrödinger cat) nuclear states.

\begin{acknowledgments}
{\it Acknowledgements:} H.E.D. was supported by EPSRC doctoral training grants. E.A.C. was supported by a Royal Society University Research Fellowship. G.G. and E.A.C. were supported by EPSRC award EP/V048333/1. A.R. acknowledges support of the Austrian Science Fund (FWF) via the Research Group FG5, I 4320, I 4380, I 3762, the Linz Institute of Technology (LIT), and the LIT Secure and Correct Systems Lab, supported by the State of Upper Austria, the European Union's Horizon 2020 research and innovation program under Grant Agreements No. 899814 (Qurope), No. 871130 (Ascent+), the QuantERA II project QD-E-QKD and the FFG (grant No. 891366). {\it Author contributions:} S.M., S.F.C.S. and A.R. developed, grew and processed the quantum dot samples. H.E.D, and G.G. conducted the experiments. H.E.D., G.G. and E.A.C. analysed the data. H.E.D., E.A.C. and G.G. drafted the manuscript with input from all authors. H.E.D. and G.G. contributed equally to this work. E.A.C. performed numerical modelling and coordinated the project.
\end{acknowledgments}


\newcommand{\RedText}[1]{{#1}}
\renewcommand{\thesection}{Supplementary Section \arabic{section}}
\setcounter{section}{0}
\renewcommand{\thefigure}{\arabic{figure}}
\renewcommand{\figurename}{Supplementary Figure}
\renewcommand{\theequation}{S\arabic{equation}}
\renewcommand{\thetable}{\arabic{table}}
\renewcommand{\tablename}{Supplementary Table}

\makeatletter
\def\l@subsection#1#2{}
\def\l@subsubsection#1#2{}
\makeatother

\pagebreak \pagenumbering{arabic}
\newpage


\section*{Supplementary Information}

\section{Sample structure}
\label{sec:Sample}

The sample structure used in this work is the same semiconductor wafer that was used previously in Refs.~\cite{MillingtonHotze2022,Zaporski2022,MillingtonHotze2023}. The sample is grown using molecular beam epitaxy (MBE) on a
semi-insulating GaAs (001) substrate. The layer sequence of the semiconductor structure is shown in Supplementary Fig.~\ref{Fig:SSample}. The growth starts with a
layer of Al$_{0.95}$Ga$_{0.05}$As followed by a single pair of
Al$_{0.2}$Ga$_{0.8}$As and Al$_{0.95}$Ga$_{0.05}$As layers acting
as a Bragg reflector in optical experiments. Then, a 95~nm thick
layer of Al$_{0.15}$Ga$_{0.85}$As is grown, followed by a 95~nm thick layer of Al$_{0.15}$Ga$_{0.85}$As
doped with Si at a volume concentration of $1.0\times10^{18}$~cm$^{-3}$. The low Al concentration of $0.15$
in the Si doped layer mitigates the issues caused by the deep DX
centers \cite{Oshiyama1986,Mooney1990,Zhai2020}. The $n$-type doped layer is followed by the
electron tunnel barrier layers: first a 5~nm thick
Al$_{0.15}$Ga$_{0.85}$As layer is grown at a reduced temperature of $560$~$^{\circ}$C to suppress Si segregation, followed by a 10~nm thick
Al$_{0.15}$Ga$_{0.85}$As and then a 15~nm thick
Al$_{0.33}$Ga$_{0.67}$As layer grown at $600$~$^{\circ}$C. Aluminium droplets are grown on the surface of the Al$_{0.33}$Ga$_{0.67}$As layer and are used to etch the nanoholes \cite{Heyn2009,Atkinson2012}. Atomic force
microscopy shows that typical nanoholes have a depth of $\approx6.5$~nm and are
$\approx70$~nm in diameter \cite{MillingtonHotze2022}. Next, a 2.1~nm thick layer of GaAs is
grown to form QDs by infilling the nanoholes as well as to form
the quantum well (QW) layer. Thus, the maximum height of the QDs
in the growth $z$ direction is $\approx9$~nm. The GaAs layer is followed by a
268~nm thick Al$_{0.33}$Ga$_{0.67}$As barrier layer. Finally, the
$p$-type contact layers doped with C are grown: a 65~nm thick
layer of Al$_{0.15}$Ga$_{0.85}$As with a
$5\times10^{18}$~cm$^{-3}$ doping concentration, followed by a 5~nm thick layer of Al$_{0.15}$Ga$_{0.85}$As with a $9\times10^{18}$~cm$^{-3}$ concentration, and a 10~nm thick layer of GaAs with a $9\times10^{18}$~cm$^{-3}$ concentration.

\begin{figure}
\centering
\includegraphics[width=0.53\linewidth]{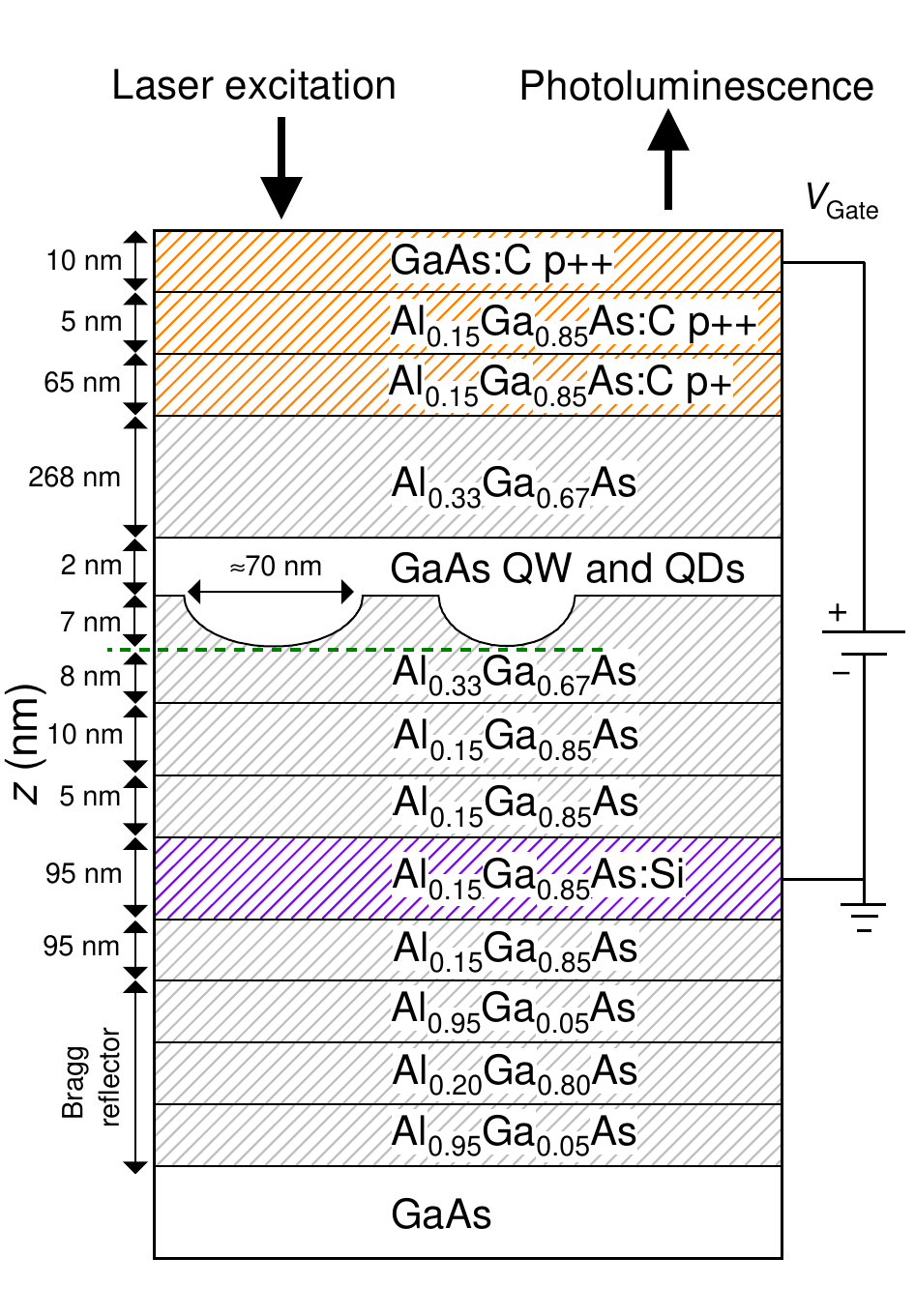}
\caption{\label{Fig:SSample} {Schematic of the quantum dot sample structure.} }
\end{figure}

The sample is processed into a $p$-$i$-$n$ diode structure. Mesa
structures with a height of 250~nm are formed by etching away the
$p$-doped layers and depositing Ni(10 nm)/AuGe(150 nm)/Ni(40 nm)/Au(100 nm) on the etched areas. The sample is then annealed to
enable diffusion down to the $n$-doped layer to form the ohmic
back contact. The top gate contact is formed by depositing Ti(15~nm)/Au(100~nm) on to the $p$-type surface of the mesa areas. Quantum dot photoluminescence (PL) is excited and collected through the top of the sample. The
sample gate bias $V_{\rm{Gate}}$ is the bias of the $p$-type top
contact with respect to the grounded $n$-type back contact. Due to the
large thickness of the top Al$_{0.33}$Ga$_{0.67}$As layer, the tunneling of holes is suppressed, whereas tunnel
coupling to the $n$-type layer enables deterministic charging of
the quantum dots with electrons by changing $V_{\rm{Gate}}$. 

In order to resolve the quadrupolar components of the nuclear magnetic resonance (NMR) spectra, the semiconductor sample is subject to a uniaxial mechanical stress. To this end, the semiconductor wafer is first cleaved into a small piece with a rectangular surface area of 0.7~mm $\times$ 2.35~mm. The edges of the rectangular profile are aligned along the $[110]$ and $[1\bar{1}0]$ crystallographic directions. The thickness of the sample along the $[001]$ growth direction is 0.35~mm. Thus, the sample is shaped as a parallelepiped. The sample is then inserted into a home-made stress cell. This is done in such a way that the two 0.7~mm $\times$ 0.35~mm surfaces of the sample are contacted to the flat titanium surfaces of the stress cell bracket. A titanium screw is then used to apply compressive stress, directed along the 2.35~mm long edge of the sample.

\section{Electron-nuclear spin system}
\label{sec:NMR}

The Hamiltonian describing the nuclear spin system includes the magnetic dipole (Zeeman) and the electric quadrupolar terms. We also consider the magnetic dipole-dipole interactions between the nuclei. The Zeeman term accounts for the coupling of the QD nuclear spins ${\bf{I}}_k$ to the static magnetic field $B_{\rm{z}}$ directed along the $z$ axis:
\begin{align}
    \mathcal{H}_{\rm{Z,N}} = -\sum_{k=1}^{N} \hbar\gamma_k B_{\rm{z}}\hat{I}_{{\rm{z}},k},\label{Eq:HZN}
\end{align}
where the summation goes over all individual nuclei $1\leq k \leq N$, $\hbar=h/(2\pi)$ is the reduced Planck's constant, $\gamma_{k}$ is the gyromagnetic ratio of the $k$-th nuclear spin and $\hat{\bf{I}}_k$ is a vector of spin operators with Cartesian components $(\hat{I}_{{\rm{x}},k},\hat{I}_{{\rm{y}},k},\hat{I}_{{\rm{z}},k})$. The result of the Zeeman term alone is a spectrum of equidistant single-spin eigenenergies $-I_{\rm{z}} \hbar \gamma_{k}B_{\rm{z}}$. These $2I+1$ states are also the eigenstates of the $\hat{I}_{{\rm{z}}}$ spin projection operator with eigenvalues $I_{\rm{z}}$ satisfying $-I\leq I_{\rm{z}}\leq +I$. 

\begin{figure}
\includegraphics[width=0.98\linewidth]{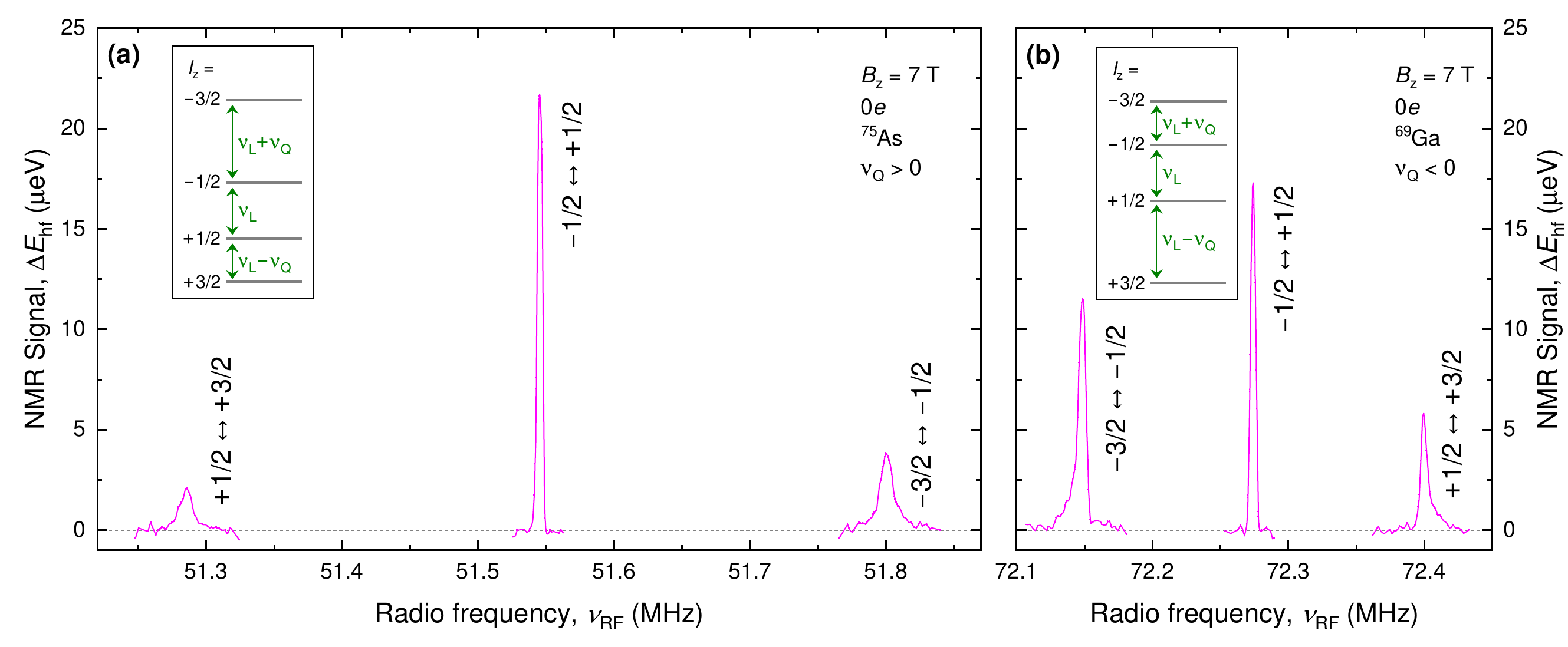}
\caption{\label{Fig:SNMR} {Nuclear magnetic resonance spectra of a single GaAs/AlGaAs quantum dot. (a) $^{75}$As spectrum. (b) $^{69}$Ga spectrum. All spectra are measured with the inverse NMR technique \cite{Chekhovich2012} at $B_{\rm{z}}\approx7$~T in a neutral QD state (0$e$). Insets show the energy level diagrams of the spin-3/2 nuclei in strong external magnetic field $B_{\rm{z}}$.}}
\end{figure}

The interaction of the nuclear electric quadrupolar moment
with the electric field gradients is described by the term (Ch.~10
in Ref.~\cite{SlichterBook}):
\begin{align}
\mathcal{H}_{\rm{Q,N}} = \sum_{k=1}^N
\frac{q_{k}}{6}[3\hat{I}_{{\rm{z'}},k}^2-I_{k}^2+\eta_{k}(\hat{I}_{{\rm{x'}},k}^2-\hat{I}_{{\rm{y'}},k}^2)],\label{Eq:HQN}
\end{align}
where $q_{k}$ and $\eta_{k}$ describe the magnitude and asymmetry
of the electric field gradient tensor, whose principal axes are
$x'y'z'$. The strain is inhomogeneous
within the QD volume, so that $q_{k}$ and $\eta_{k}$ vary
between the individual nuclei. The axes $x'y'z'$ are different for
each nucleus and generally do not coincide with crystallographic
axes or magnetic field direction. For the as-grown GaAs/AlGaAs QDs the quadrupolar shifts are around $\vert q_{k} \vert/h\approx20$~kHz \cite{Ulhaq2016}, reaching $q_{k}/h\approx200$~kHz for a small fraction of the nuclei \cite{Zaporski2022}. In the studied structure, $q_{k}$ are dominated by the extrinsic uniaxial stress. All experiments are conducted under sufficiently strong magnetic fields, where $|\hbar\gamma_k B_{\rm{z}}|\gg |q_{k}|$ and quadrupolar effects can be treated perturbatively. In this perturbative regime, the main effect of the quadrupolar shifts is the anharmonicity of the nuclear spin eigenenergies and the resulting quadrupolar NMR multiplet of $2I$ magnetic-dipole transitions, split by $\nu_{\rm{Q}}\approx q_{k}/h$. The $I_{\rm{z}}=\pm1/2$ states of a half-integer nuclear spin are influenced by quadrupolar effects only in the second order. These second order shifts scale as $\propto\nu_{\rm{Q}}^2/\nu_{\rm{N}}$, where $\nu_{\rm{N}}=\gamma B_{\rm{z}}/(2\pi)$ is the nuclear spin Larmor frequency.

The nuclear spin energy spectrum is probed using optically-detected NMR spectroscopy of individual QDs. Supplementary Fig.~\ref{Fig:SNMR} shows the NMR spectra of $^{75}$As (a) and $^{69}$Ga (b) spin-3/2 nuclei measured on an empty (0$e$) QD. The insets show the corresponding diagrams of the nuclear spin energy levels and the allowed magnetic dipole NMR transitions. As expected for $I=3/2$, each NMR spectrum is a triplet. The first order quadrupolar shifts are $\nu_{\rm{Q}}\approx+260$~kHz and $\nu_{\rm{Q}}\approx-125$~kHz for $^{75}$As and $^{69}$Ga, respectively. The signs of $\nu_{\rm{Q}}$ are opposite due to the opposite signs of the gradient elastic tensors of the group-III and group-V elements \cite{Chekhovich2018}. The quadrupolar shifts are the witnesses of the strain induced by the external stress. The strain is estimated to be $\epsilon_{\rm{b}}\approx0.0025$.

Direct interaction between the nuclei is described by the dipole-dipole Hamiltonian:
\begin{align}
\mathcal{H}_{\rm{DD}}=\sum_{1\leq j<k\leq N}b_{j,k}\left(3\hat{I}_{{\rm{z}},j}\hat{I}_{{\rm{z}},k}-\hat{\bf{I}}_j{\bf{\cdot}}\hat{\bf{I}}_k\right){\rm{,}}\nonumber\\
b_{j,k}=\frac{\mu_0 \hbar^2}{4\pi}\frac{\gamma_{j}\gamma_{k}}{2}\frac{1-3\cos^2{\theta_{j,k}}}{r_{j,k}^3}\label{Eq:HDD}
\end{align}
Here, $\mu_0=4\pi\times 10^{-7}\;{\rm{N A}}^{-2}$ is the magnetic constant and $r_{j,k}$ denotes the length of the vector, which forms an angle $\theta$ with the $z$ axis and connects the two spins $j$ and $k$. The Hamiltonian of Supplementary Eq.~(\ref{Eq:HDD}) has been truncated to eliminate all spin non-conserving terms -- this is justified for static magnetic field exceeding $\gtrsim1$~mT. The typical magnitude of the interaction constants for the nearby nuclei in GaAs is $\max{(|b_{j,k}|)}/h\approx100$~Hz.  Consequently, the typical timescales of the processes driven by the many-body dipole-dipole interactions are on the order of 1~ms. This is much slower than the duration of the QND measurement $T_{\rm{RF}}\approx10~\mu$s. As a result, nuclear-nuclear interactions play only a minor role in the context of the current work.

The interaction of the conduction band electron spin $\bf{s}$ with
the ensemble of the QD nuclear spins is dominated by the contact
(Fermi) hyperfine interaction, with the following Hamiltonian:
\begin{align}
\mathcal{H}_{\rm{hf}}=\sum_{k=1}^N{a_k(\hat{s}_{\rm{x}}\hat{I}_{{\rm{x}},k}+\hat{s}_{\rm{y}}\hat{I}_{{\rm{y}},k}+\hat{s}_{\rm{z}}\hat{I}_{{\rm{z}},k})},\label{Eq:Hhfe}
\end{align}
where the hyperfine constant of an individual nucleus $k$ is
$a_k=A^{(k)}|\psi({\bf{r}}_k)|^2{\it{v}}$. Unlike $a_k$, the hyperfine constant $A^{(k)}$ is a parameter describing only the
material and the isotope type to which nucleus $k$ belongs,
$|\psi({\bf{r}}_k)|^2$ is the density of the electron envelope
wavefunction at the nuclear site ${\bf{r}}_k$ of the
crystal lattice, and ${\it{v}}$ is the crystal volume per one
cation or one anion. 

The definitions of the hyperfine constants differ between different sources. With the definition adopted here, a fully polarized isotope with spin $I$, hyperfine constant $A$ and a 100\% abundance (e.g. $^{75}$As), would shift the energies of the electron spin states $s_{\rm{z}}=\pm1/2$ by $\pm AI/2$, irrespective of the shape of $|\psi({\bf{r}})|^2$. With such definition, the typical values in GaAs are $A\approx50$~$\mu$eV (Ref.~\cite{Chekhovich2017}). The frequency Knight shift of an individual nucleus coupled to a spin polarized electron $s_{\rm{z}}=\pm1/2$ is $\pm a_k/(2h)$. The typical Knight shift can be estimated from the frequency-detuned single-shot NMR spectra of Fig.~2(b) of the main text. We find $a_j/(h)\approx140$~kHz or $a_j\approx0.58$~neV for $^{69}$Ga isotope. Taking the ratio $A/a_j$ we roughly estimate the effective number of nuclei $N\approx10^5$ coupled to the QD electron spin.

The hyperfine interaction of the valence band holes is an order of magnitude smaller \cite{Chekhovich2013NPhys} and can be ignored in the context of this work.

\section{Experimental details and additional results}
\label{sec:ExpTechn}

The sample is placed in a liquid helium bath cryostat. A
superconducting coil is used to apply magnetic field up to
$B_{\rm{z}}=8$~T. The field is parallel to the sample growth
direction $[001]$ and the optical axis $z$ (Faraday geometry). The field and the optical axis are orthogonal to the direction of the applied mechanical stress. We use a confocal microscopy configuration. An aspheric lens with a focal
distance of 1.45~mm and NA = 0.58 is used as an objective for
optical excitation of the QD and for photoluminescence (PL)
collection. The excitation laser is focused into a spot
with a diameter of $\approx1~\mu$m. The collected PL is dispersed
in a two-stage grating spectrometer, each stage with a 1~m
focal length, followed by a pair of achromatic lens doublets, which transfers the spectral image onto a charge-coupled device (CCD) photo-detector with a magnification of 3.75. The changes in the spectral splitting $\Delta E_{\rm{PL}}$ of either a neutral exciton $X^0$ or a negatively
charged trion $X^-$, derived from the PL spectra, are used to
measure the hyperfine shifts $E_{\rm{hf}}$ proportional to the
nuclear spin polarization degree.

\begin{figure}
\includegraphics[width=0.98\linewidth]{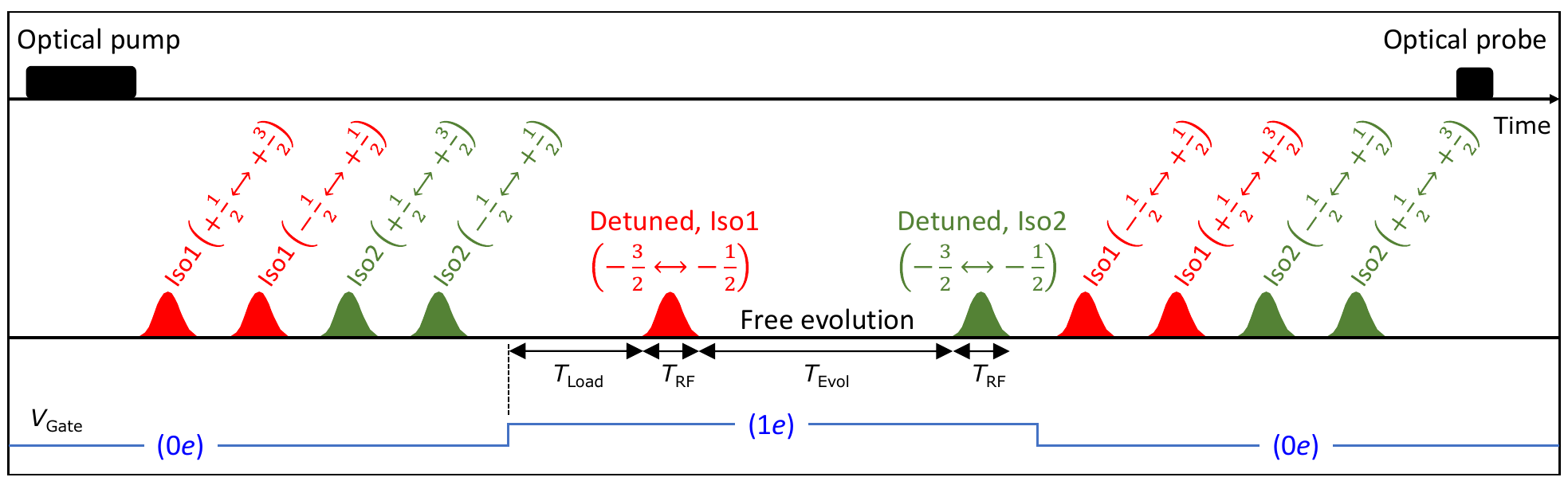}
\caption{\label{Fig:STDiag} {Timing diagram of the pulsed nuclear magnetic resonance experiment.}}
\end{figure}

Supplementary Fig.~\ref{Fig:STDiag} is a detailed version of Fig.~1(e) of the main text and shows the timing of a pulsed single-shot NMR measurement. In what follows we describe the individual steps of the timing sequence. 

\subsection{Optical pumping of the quantum dot nuclear spins}

Optical pumping of the QD nuclear spin polarization (labelled Pump in Supplementary Fig.~\ref{Fig:STDiag}) is achieved using the emission of a tunable single-mode circularly polarized diode laser. Optical dynamical nuclear spin polarization is a well known process, that has been observed in many types of QDs \citep{Gammon2001,Eble2006,Skiba2008,Ulhaq2016,Ragunathan2019}, see Ref. \citep{Urbaszek2013} for a review. In the context of the present study, we simply rely on the fact that optical pumping is a reliable tool for achieving nuclear spin polarizations exceeding $50\%$ on a timescales of a few seconds. The physics of nuclear spin pumping in the semiconductor wafer studied here are discussed in Ref. \citep{MillingtonHotze2023}. In brief, dynamic nuclear polarization is a three-stage cyclic process. At the first stage a spin polarized electron is created optically. This is made possible by the selection rules, which allow conversion of the circularly polarized photons into spin-polarized electron-hole pairs in group III-V semiconductors. At the second stage, the electron exchanges its spin with one of the nuclei through the flip-flop term of the electron-nuclear hyperfine Hamiltonian (Supplementary Eq.~\ref{Eq:Hhfe}). The third stage is the electron-hole optical recombination or tunnel escape, which removes the flipped electron. This final step is required in order to let the QD accept new spin-polarized electrons and continue polarizing the ensemble of $N\approx 10^5$ nuclear spins of the QD. During the optical pump the sample gate is set to a large reverse bias, typically $V_{\rm{Gate}}=-2$~V. The pump power is $\approx1$~mW, which is three orders of magnitude higher than the ground-state PL saturation power. The photon energy of the pump laser is typically $\approx5-10$~meV above the $X^0$ PL energy. The pump duration is $T_{\rm{Pump}}=3.5$~s.

\subsection{Nuclear magnetic resonance}

The oscillating magnetic field $B_{\rm{x}}\perp z$, required to perform NMR, is produced by a copper wire coil placed at a distance of $\approx0.5$~mm
from the QD sample. The coil is made of 10 turns of a 0.1~mm diameter enameled copper wire wound on a $\approx0.4$~mm diameter spool in 5 layers, with 2 turns in each layer. The coil is driven by a class-AB RF amplifier (Tomco BT01000-AlphaSA rated up to 1000~W) which is fed by the output of an arbitrary waveform generator Keysight M8190.

Supplementary Fig.~\ref{Fig:STDiag} shows the timing diagram of a two-pulse experiment, which we now discuss in more detail. We consider the case of $\sigma^+$ polarized optical pump, which produces a Boltzmann distribution of the nuclear spin $z$ projections \cite{MillingtonHotze2023}, populating predominantly the $I_{z}=-3/2$ nuclear spin state, while leaving the $I_{z}=+3/2$ state the least populated. The electron to nuclear spin state transfer is performed using the $-3/2\leftrightarrow-1/2$ NMR transition. In order to increase the NMR signal, we maximize the initial population difference of the $I_{z}=-3/2,-1/2$ nuclear states. This is achieved through state population transfer \cite{Chekhovich2015}. A pair of RF pulses performing $\pi$ rotation (i.e. inversion) is applied to each of the two isotopes used in the experiment. First, the $+1/2\leftrightarrow+3/2$ transition is driven to exchange the populations of the $I_{z}=+1/2,+3/2$ states of the first isotope (labelled ``Iso1'' in Supplementary Fig.~\ref{Fig:STDiag}), making $I_{z}=+1/2$ the least populated state. The second $\pi$ pulse applied to the $-1/2\leftrightarrow+1/2$ transition transfers the smallest population to $I_{z}=-1/2$. The same population-transfer sequence is applied to the second isotope (labelled ``Iso2'' in Supplementary Fig.~\ref{Fig:STDiag}). The state transfer is performed under reverse bias, which keeps the QD empty of all charges (0$e$). The absence of charges eliminates the Knight shifts, thus maximizing the fidelity of the state transfer performed by the NMR pulses.

Following the NMR state transfer, the sample gate bias is increased in order to load a single electron from the Fermi reservoir into the QD (1$e$ charge state). The actual electron tunneling takes place on a submicrosecond timescale. However, the electron is then left to equilibrate for a time interval $T_{\rm{Load}}$. We typically use $T_{\rm{Load}}\in[30, 90]$~ms. This is much longer than the measured electron spin lifetimes $T_{1,{\rm{e}}}$, ensuring that any transient effects decay before the electron spin state is measured. 

Following the electron equilibration, a frequency-detuned RF $\pi$ pulse is applied to the $-3/2\leftrightarrow-1/2$ transition of the first isotope (``Iso1''). The detuning is chosen to be close to the weighted average nuclear hyperfine (Knight) shift and is typically $\approx70$~kHz for $^{69}$Ga in the studied QDs. This pulse performs a QND measurement of the electron spin, storing the outcome in the long-lived longitudinal nuclear spin polarization of ``Iso1''. The electron is then left for a time $T_{\rm{Evol}}$ to evolve freely without any optical or RF excitation. This is followed by the second frequency-detuned RF $\pi$ pulse applied to ``Iso2'' in order to perform the second QND measurement of the electron spin. The pulses have cosine (near-Gaussian) envelopes and a total duration $T_{\rm{RF}}$, counted between the zero-amplitude points at the start and the end of the pulse (the corresponding full width at half maximum is $T_{\rm{RF}}/2$). Depending on the measurement,  $T_{\rm{RF}}$ is chosen to be between 10 and 40~$\mu$s.

The results of the QND measurements are encoded in the $I_{z}=-3/2,-1/2$ nuclear spin subspaces where either the $I_{z}=-3/2$ or the $I_{z}=-1/2$ states are more populated, depending on the electron spin projection state $s_{\rm{z}}$. At this stage, the nuclear spin polarization can already be retrieved optically. However, it is beneficial to multiply the NMR signal, by exploiting the entire $I=3/2$ Hilbert space. To this end, the sample is biased back into an empty-QD state (0$e$) and the reverse population transfer is performed on both isotopes. Each reverse population transfer consists of a $\pi$ pulse on $-1/2\leftrightarrow+1/2$, followed by a $\pi$ pulse on the $+1/2\leftrightarrow+3/2$ NMR transition. This provides a factor of $\approx3$ amplification in the NMR signal, which can be understood as follows. If the QND frequency-detuned pulse leaves the $I_{z}=-3/2$ state as the most populated, then the reverse population transfer has only a small effect on the nuclear spin polarization. In the opposite case, if QND leaves the $I_{z}=-1/2$ state as the most populated, the reverse population transfer makes the $I_{z}=+3/2$ state the most populated. Thus, instead of the $I_{z}=-3/2,-1/2$ subspace, the reverse population transfer encodes the NMR signal in the $I_{z}=-3/2,+3/2$ subspace, which approximately triples the optically-detected hyperfine shift $\Delta E_{\rm{hf}}$.

In some cases it is more convenient to perform QND using the $I_{z}=+1/2,+3/2$ nuclear spin subspaces, or a combination, where the $I_{z}=+1/2,+3/2$ states are used on one of the isotopes, while the $I_{z}=-3/2,-1/2$ states are used for the other isotope. In such cases the initial and reverse population transfer sequences are altered to match the chosen transitions, however, the principle described above remains the same. For the two-pulse experiments shown in Fig.~3(e) of the main text, the first RF pulse is detuned to the higher frequencies from the $-3/2\leftrightarrow-1/2$ high-frequency satellite of $^{75}$As, while the second pulse is detuned to the lower frequencies from the $-3/2\leftrightarrow-1/2$ low-frequency satellite of $^{69}$Ga. In those experiments where only one RF pulse is used on a single isotope, the timing diagram is the same as in Supplementary Fig.~\ref{Fig:STDiag}, but omitting the pulses for the second isotope.

The number of the nuclei used in the QND measurement can be estimated as follows. The root mean square number of nuclei $N \approx 6.5 \times 10^4$ has been estimated previously from the electron spin dephasing dynamics in the same QD sample \cite{Zaporski2022}. Half of these nuclei are arsenic. Moreover, due to the incomplete polarization of the nuclei, only $\approx80\%$ are in the two-level subspace (such as $I_{z}=-3/2,-1/2$) used for the electron spin readout. Therefore, when $^{75}$As is used for the readout, the number of active nuclei is $ \approx \frac{1}{2} \times 0.8 \times 6.5 \times 10^4 \approx 2.6 \times 10^4$. When the readout is carried out via $^{69}$Ga, we need to take into account the $\approx60\%$ natural abundance of the isotope, which leads to $\approx 1.6 \times 10^4$ nuclei actively used in the electron spin QND measurement.

All NMR measurements are differential. In addition to the actual single-shot measurements with the timing shown in Supplementary Fig.~\ref{Fig:STDiag}, we periodically collect the reference PL probe spectra. These reference probe spectra are measured with exactly the same RF pulse sequence, including population transfers, but without the detuned RF pulses. The difference of the spectral splittings from the two measurements yields the differential NMR signal $\Delta E_{\rm{hf}}$.

\subsection{Optical probing of the quantum dot nuclear spins}

For optical probing of the nuclear spin polarization we use a diode laser emitting at 760~nm. Sample forward bias, typically $V_{\rm{Gate}}=+0.9$~V, and the probe power are chosen to maximize (saturate) PL intensity of the ground state neutral exciton $X^0$. Fig.~2(a) of the main text shows neutral exciton ($X^0$) PL probe spectra measured at $B_{\rm{z}}=5.31$~T following optical pumping with a $\sigma^-$ circularly polarized laser. The PL arises from recombination of the electron-hole pairs in a QD (neutral exciton $X^0$). The PL spectrum is a doublet, where each component corresponds to the optically-excited electron in a spin-up ($s_{\rm{z}}=+1/2$) or spin-down ($s_{\rm{z}}=+1/2$) projection state. Consequently, the splitting $\Delta E_{\rm{PL}}$ of the PL spectral doublet is sensitive to polarization of the nuclei along the $z$ axis. The variation of this spectral splitting reveals the variation of the hyperfine shift $\Delta E_{\rm{hf}}$. These shifts are used to monitor the average QD nuclear spin polarization in NMR experiments.  It is also possible to probe the nuclear spin polarization using PL of a negatively charged trion $X^-$. The charge state for optical probing is selected for each individual QD and magnetic field strength. This choice is governed by the linewidths and the brightness of PL.

\begin{figure}
\includegraphics[width=0.4\linewidth]{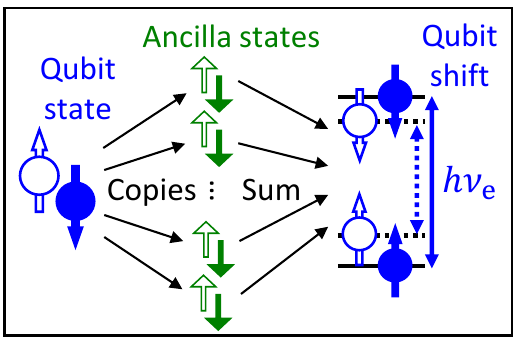}
\caption{\label{Fig:SCopyToSum} Schematic of the qubit readout concept. First, a detuned radiofrequency pulse copies the electron spin qubit into a large number of nuclear spin ancillae. Next, the arithmetic sum of the ancilla states is read out by measuring the hyperfine shifts of the electron spin energy gap $h \nu_{\rm{e}}$. These energy shifts are probed through repeated optical injection of electrons into the QD.}
\end{figure}

Illumination with a probe laser inevitably acts back on the nuclear spin polarization. Each optically excited electron has a finite probability to flip one of $N\approx10^5$ nuclear spins. Eventually, the nuclear spin polarization reaches its steady-state, governed only by the power, wavelength and polarization of the probe laser. Therefore, the duration of the probe $T_{\rm{Probe}}$ must be short enough to retain sufficient information about the nuclear polarization at the start of the probe. On the other hand, at the optical probing stage we are only interested in the average nuclear spin polarization, i.e. the arithmetic sum of all the copies transferred by the detuned RF pulse from the electron spin qubit into the nuclear spin $z$ projections (see Supplementary Fig.~\ref{Fig:SCopyToSum}). In other words, optical probing measures the $z$ projection of a large total spin $\approx N I$ formed by thousands of nuclei. Due to the large $N$, this total spin is essentially a classical variable. Consequently, the QD can be excited optically many times, and a large number of photons can be collected before the decay of the average nuclear spin polarization causes any significant distortion in the measurement outcome. In order to determine the optimal $T_{\rm{Probe}}$ we perform calibration measurements, with a typical result shown in Supplementary Fig.~\ref{Fig:SProbe}. In this experiment the QD is first pumped with a $\sigma^+$ or $\sigma^-$ polarized laser in order to create large initial nuclear polarization. A probe laser pulse is then applied. The PL spectral splitting $\Delta E_{\rm{PL}}$ is measured at the end of this probe. It can be seen that the probe induces decay of the nuclear spin polarization on a timescale of a few hundred milliseconds. The probe time $T_{\rm{Probe}}$ used in the single-shot NMR experiments is chosen to ensure that any distortion of the measured $\Delta  E_{\rm{hf}}$ is small enough to resolve the $\Delta  E_{\rm{hf}}$ values corresponding to the opposite electron spin states. For example, for the data shown in Supplementary Fig.~\ref{Fig:SProbe} we choose $T_{\rm{Probe}}=0.05$~s.

\begin{figure}
\includegraphics[width=0.6\linewidth]{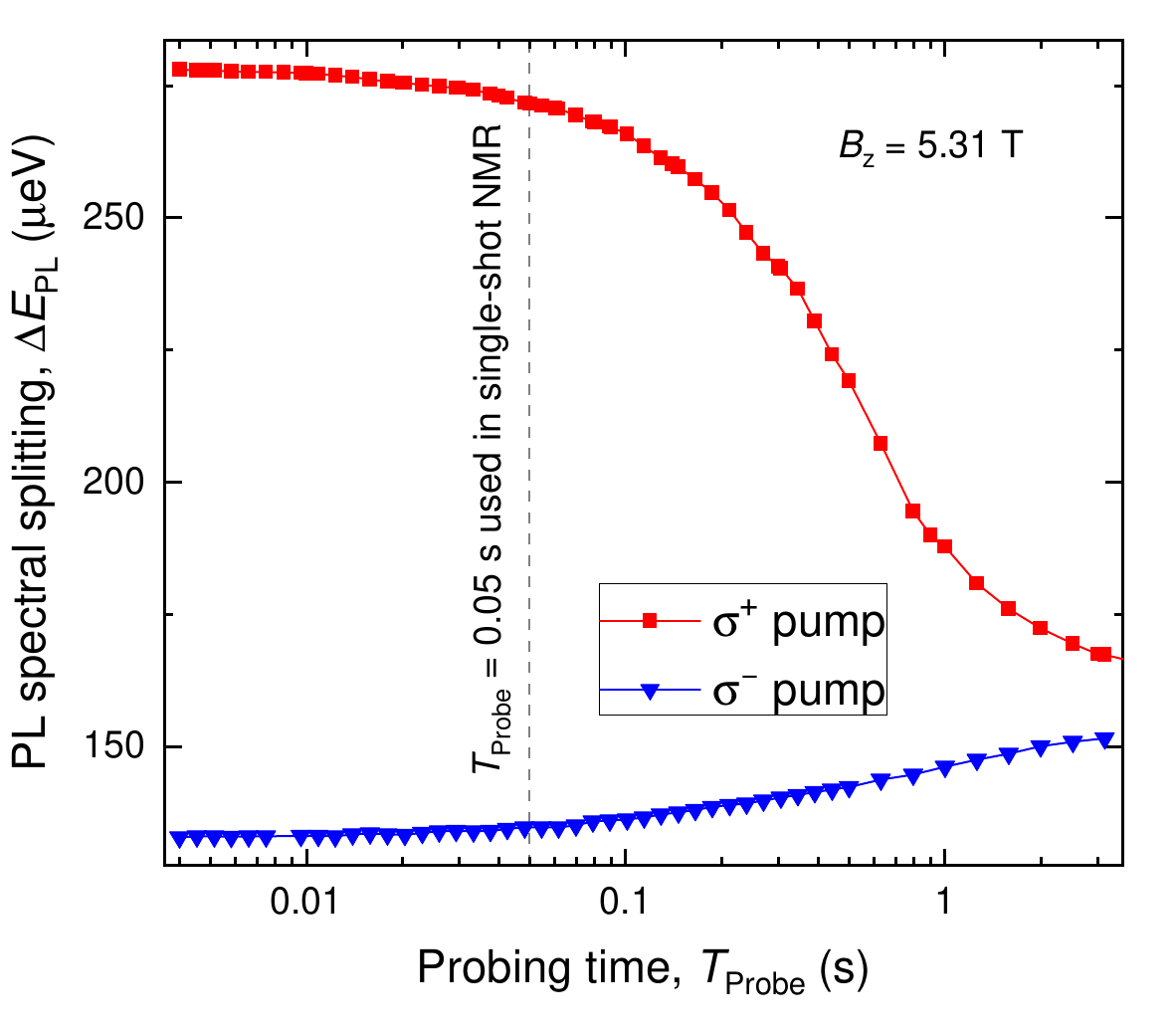}
\caption{\label{Fig:SProbe} Calibration of the optical probing of the QD nuclear spin polarization. Hyperfine shift measured as a function of the probing time $T_{\rm{Probe}}$ following a $\sigma^+$ or $\sigma^-$ pumping of the nuclear spin polarization in a QD. Vertical dashed line shows the $T_{\rm{Probe}}$ value chosen for the NMR measurements on this individual QD at this particular magnetic field of $B_{\rm{z}}=5.31$~T.}
\end{figure}

We note that due to the weak backaction it is possible in principle to apply multiple optical probe pulse. This could be useful for isotope-selective retrieval of the NMR signals. For example, one can use a (Probe - RF $\pi$ pulse - Probe) sequence to measure selectively the NMR transition chosen by the RF pulse. This could be beneficial in situations such as those shown in Supplementary Figs.~\ref{Fig:SHist}(d)--\ref{Fig:SHist}(f) of the main text, where the ``no-flip'' NMR signals arising from $s_{\rm{z}}=\pm1/2$ overlap. However, in the present work we rely on a simple implementation with one probe pulse, which is sufficient to demonstrate the concept.

\section{Readout fidelity modelling}
\label{sec:FFit}

Here we discuss how the fidelity of the spin readout is derived from the histograms, such as shown in Figs.~3(b) and 3(c) of the main text. We need to construct the model probability distribution of the single-shot NMR signal amplitudes $\Delta E_{\rm{hf}}$. When constructing the probability density, we take into account the finite probability for the electron spin to flip during the RF pulse duration $T_{\rm{RF}}$. The flips are modelled as a random telegraph process, where the flips are instantaneous (infinitely fast) and are not correlated. In principle, the electron spin can flip $m$ times during $T_{\rm{RF}}$, with the probability of such events scaling as $\propto (T_{\rm{RF}}/T_{\rm{1,e}})^m$, where $T_{\rm{1,e}}$ is the electron spin lifetime. In all our experiments $T_{\rm{RF}}\ll T_{\rm{1,e}}$, so we only account for the terms up to the first order $m\leq1$ (in other words, we ignore the possibility for the electron to flip more than once during $T_{\rm{RF}}$).

For a telegraph process, the probability of zero flips occurring during any given time interval $T_{\rm{RF}}$ is:
\begin{eqnarray}
p_{m=0} = \exp\left[{-\frac{\left(1-\rho_{\rm{e}} ^2\right) T_{\rm{RF}}}{2 T_{\rm{1,e}}}}\right],\label{Eq:pNoFlip}
\end{eqnarray}
where $-1\leq \rho_{\rm{e}} \leq 1$ is the equilibrium electron spin polarization degree. For the probability to have exactly one electron spin flip we write:
\begin{eqnarray}
p_{m=1} = 1-p_{m=0},\label{Eq:pOneFlip}
\end{eqnarray}
where we have used the simplifying assumption that two or more flips ($m\geq2$) are not possible.

In the case of zero flips ($m=0$), the electron is either in a spin up ($s_{{\rm{z}}}=+1/2$) or spin down ($s_{{\rm{z}}}=-1/2$) state throughout the entire RF measurement pulse. In the absence of noise, these two spin states will result in two discrete NMR readout values: $\Delta E_{\rm{hf}}^+$ and $\Delta E_{\rm{hf}}^-$, respectively. The relevant probability distributions are described by the delta functions centered at $\Delta E_{\rm{hf}}^+$ and $\Delta E_{\rm{hf}}^-$ for $s_{{\rm{z}}}=+1/2$ and $-1/2$. The total probabilities of detecting $s_{{\rm{z}}}=+1/2$ and $-1/2$ in equilibrium are $(1+\rho_{\rm{e}})p_{m=0}/2$ and $(1-\rho_{\rm{e}})p_{m=0}/2$, respectively.

\begin{figure*}
\includegraphics[width=0.7\linewidth]{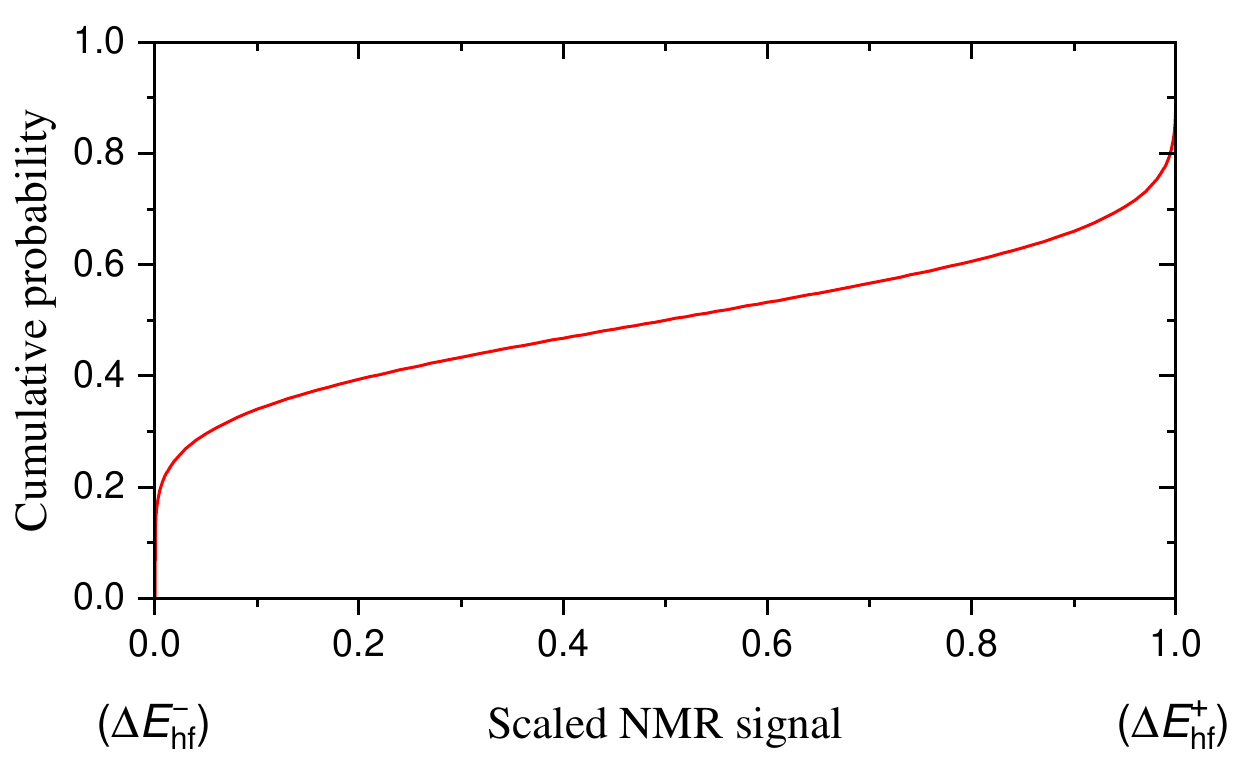}
\caption{Cumulative distribution function of the NMR signals detected in case where the electron spin flips once during the measurement RF pulse. The amplitude of the RF pulse is assumed to have a cosine envelope. The horizontal axis is scaled in the range $[\Delta E_{\rm{hf}}^-,\Delta E_{\rm{hf}}^+]$.} \label{Fig:CDF}
\end{figure*}

Next, we consider the $m=1$ case, where the electron flips at a timepoint $t_{\rm{Flip}}$. The distribution of $t_{\rm{Flip}}$ is uniform in the interval $[0,T_{\rm{RF}}]$. In order to evaluate the effect of the electron spin flip on the measured NMR signal we use simple geometrical calculations. We assume that all nuclear spins are exactly on resonance (completely out of resonance) with the RF pulse when the electron is in the $s_{{\rm{z}}}=+1/2$ ($s_{{\rm{z}}}=-1/2$) spin state. Consider the case where the electron is in the $s_{{\rm{z}}}=+1/2$ state at the start of the RF pulse. Then the nuclei will be rotated from their initial $I_{{\rm{z}}}=+1/2$ states towards the inverted $I_{{\rm{z}}}=-1/2$ states. However, this rotation is interrupted at $t_{\rm{Flip}}$. We then calculate the change in $I_{{\rm{z}}}$ arising from such an interrupted rotation in case of an RF pulse with a cosine shaped envelope. Linear rescaling of this change in $I_{{\rm{z}}}$ yields the NMR signal as a function of $t_{\rm{Flip}}$:
\begin{eqnarray}
\Delta E_{\rm{hf}}(t_{\rm{Flip}})=\Delta E_{\rm{hf}}^- + (\Delta E_{\rm{hf}}^+ - \Delta E_{\rm{hf}}^-)\frac{1}{2} \left(1-\cos \left[\frac{1}{2} \sin \left(\frac{2 \pi  t_{\text{Flip}}}{T_{\text{RF}}}\right)-\frac{\pi t_{\text{Flip}}}{T_{\text{RF}}}\right]\right)\label{Eq:FlippedNMR}
\end{eqnarray}
This function is a scaled sigmoid, and, given that $t_{\rm{Flip}}$ is distributed uniformly, its appropriately scaled inverse is a cumulative distribution function (CDF) of the single-shot NMR signals $\Delta E_{\rm{hf}}$ detected in the case of exactly one electron flip ($m=1$). There is no analytical inverse. The numerically evaluated CDF is shown in Supplementary Fig.~\ref{Fig:CDF}. Its derivative is the probability density function (PDF) and also has to be evaluated numerically. Some properties of this PDF can be noted from Supplementary Fig.~\ref{Fig:CDF}. Namely, it is approximately a sum of two sharp modes at $\Delta E_{\rm{hf}}^-$ and $\Delta E_{\rm{hf}}^+$, and a relatively uniform background in between these modes. This shape can be understood qualitatively as originating from the cosine envelope of the RF pulse. If the electron flips near the start or the end of the pulse, where the RF amplitude is $\approx0$, the nuclei either have not rotated much, or have already been flipped, respectively. Thus, the flips at the start and the end of the RF pulse give rise to NMR signals very close to $\Delta E_{\rm{hf}}^-$ and $\Delta E_{\rm{hf}}^+$, respectively. The case where the electron is in the $s_{{\rm{z}}}=-1/2$ state at the start of the RF pulse gives exactly the same distribution of the NMR signals $\Delta E_{\rm{hf}}$.

Summarising, we have the following three contributions to the distribution of the NMR signals $\Delta E_{\rm{hf}}$: 
\begin{itemize}
\item (i) The delta-peak mode at $\Delta E_{\rm{hf}}^+$ with a probability $\frac{1+\rho_{\rm{e}}}{2}\exp\left[{-\frac{\left(1-\rho_{\rm{e}} ^2\right) T_{\rm{RF}}}{2 T_{\rm{1,e}}}}\right]$, corresponding to the $s_{{\rm{z}}}=+1/2$ electron state with no flips.
\item (ii) The delta-peak mode at $\Delta E_{\rm{hf}}^-$ with a probability $\frac{1-\rho_{\rm{e}}}{2}\exp\left[{-\frac{\left(1-\rho_{\rm{e}} ^2\right) T_{\rm{RF}}}{2 T_{\rm{1,e}}}}\right]$, corresponding to the $s_{{\rm{z}}}=-1/2$ electron state with no flips.
\item (iii) The bimodal distribution given by Supplementary Eq.~\ref{Eq:FlippedNMR} with a probability $1-\exp\left[{-\frac{\left(1-\rho_{\rm{e}} ^2\right) T_{\rm{RF}}}{2 T_{\rm{1,e}}}}\right]$, corresponding to the case where the electron spin flips once during the RF pulse.
\end{itemize}
These three contributions, weighted by their relevant probabilities, are added together to obtain the complete ideal PDF. The final step is to take into account the non-ideal nature of the optical readout of the single-shot NMR signals. The readout noise is modelled by assuming that the detected $\Delta E_{\rm{hf}}$ values have a Gaussian distribution with a full width at half maximum $w$ and centred on the true $\Delta E_{\rm{hf}}$. Thus, we convolve the complete ideal PDF with a Gaussian PDF. This convolved PDF is then fitted to the experimental single-shot NMR data, such as shown in the histograms of Figs.~3(b) and 3(c) of the main text. The best fit is found by maximizing the likelihood estimator, with $\Delta E_{\rm{hf}}^-$, $\Delta E_{\rm{hf}}^+$, $\rho_{\rm{e}}$, $T_{\rm{1,e}}$ and $w$ used as fitting parameters. 

The best fits are shown by the solid lines in Figs.~3(b) and 3(c) of the main text. The best fit Gaussian FWHM is $w\approx2.4-2.7$~$\mu$eV, which is a very good match to $w\approx2.4$~$\mu$eV found in the reference measurement on a neutral QD [0$e$, Fig.~3(a) of the main text] where the data is fitted with a single Gaussian. This agreement confirms that optical readout noise is the main source of the histogram mode broadening and validates the model used for the 1$e$ single-shot NMR. The fitted equilibrium electron spin polarization is small $|\rho_{\rm{e}}|\leq0.1$, owing to the small electron $g$-factor in the studied QDs \cite{Ulhaq2016,Zaporski2022}. The best fit values for the electron spin lifetimes are $T_{\rm{1,e}}\approx1.7$~ms and $T_{\rm{1,e}}\approx0.58$~ms at $B_{\rm{z}}=1.6$~T and 5.3~T, respectively. These values are somewhat smaller than the more accurate measurements $T_{\rm{1,e}}\approx8.7$~ms and $T_{\rm{1,e}}\approx0.71$~ms obtained at the same magnetic fields via direct measurement of the electron spin relaxation. The $T_{\rm{1,e}}$ parameter used in the model probability distribution essentially encodes the qubit readout infidelity $1-F$. The underestimated fitted values of $T_{\rm{1,e}}$ probably mean that the true qubit readout fidelities are even closer to unity than derived from fitting.

Once the fitting parameters are obtained, the qubit readout criterion is defined by setting the threshold value in the middle of the two modes at $(\Delta E_{\rm{hf}}^- + \Delta E_{\rm{hf}}^+)/2$. Any single-shot NMR signal $\Delta E_{\rm{hf}}$ below (above) this threshold is interpreted as $s_{{\rm{z}}}=-1/2$ ($s_{{\rm{z}}}=+1/2$). The readout fidelity $F$ is defined as the average probability to detect the true $s_{{\rm{z}}}$ correctly. We have $F=p_{m=0}p_{\rm{Opt}}+p_{m=1}/2$, where $p_{\rm{Opt}}$ denotes the probability that the optical readout noise does not cause the measured $\Delta E_{\rm{hf}}$ to cross the detection threshold. The two modes in the measured histograms are well resolved, especially at high magnetic field [Fig.~3(c) of the main text], so that $p_{\rm{Opt}}\approx1$. Therefore, the fidelity is $F=p_{m=0}p_{\rm{Opt}}+p_{m=1}/2\approx p_{m=0}+p_{m=1}/2 = 1-p_{m=1}/2$, dominated by the probability that the electron flips during the readout RF pulse. The $1/2$ factor in $p_{m=1}/2$ accounts for the fact that even when the electron spin is flipped randomly, there is still a 50\% probability that the measured $\Delta E_{\rm{hf}}$ stays on the correct side of the detection threshold and the electron spin projection  $s_{{\rm{z}}}$ is measured correctly. We find $F\approx0.9985$ at both $B_{\rm{z}}=1.6$~T and 5.3~T. We note that the same value of $F$ is found despite the longer electron spin lifetime at $B_{\rm{z}}=1.6$~T. This is likely explained by the longer measurement pulse $T_{\rm{RF}}\approx20~\mu$s, as opposed to $T_{\rm{RF}}\approx10~\mu$s used at $B_{\rm{z}}=5.3$~T, and the smaller separation of the histogram modes, which means that $p_{\rm{Opt}}$ is not as close to unity as it is at high magnetic field.

\section{Additional data}
\label{sec:AddData}

\begin{figure*}
\includegraphics[width=0.9\linewidth]{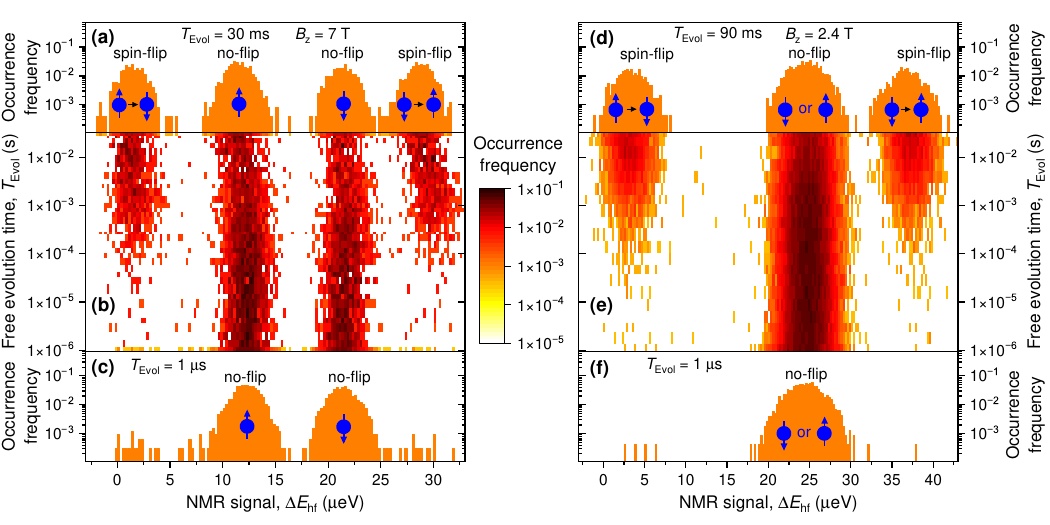}
\caption{(a-c) Histograms of the single-shot NMR signals $\Delta E_{\rm{hf}}$ measured at $B_{\rm{z}}=7$~T with two RF $\pi$ pulses applied to $^{75}$As and $^{69}$Ga and delayed by $T_{\rm{Evol}}$. A full 2D histogram at variable $T_{\rm{Evol}}$ is shown in (b), while (a) and (c) show the cross sections at long and short $T_{\rm{Evol}}$, respectively. (d-f) Same as (a-c) but on a different QD and at a lower magnetic field $B_{\rm{z}}=2.4$~T. The two ``no-flip'' modes are not resolved in this measurement.} \label{Fig:SHist}
\end{figure*}

Supplementary Figs.~\ref{Fig:SHist}(a)--\ref{Fig:SHist}(c) replicate Figs.~3(d)--3(e) of the main text, where the results are shown for an experiment with two RF pulses.
The first pulse applied to $^{75}$As nuclei records the initial state of the electron spin, while the second pulse on $^{69}$Ga stores the state after the interpulse free-evolution delay $T_{\rm{Evol}}$. The optically-measured $\Delta E_{\rm{hf}}$ is the total NMR signal produced by the two pulses. Supplementary Fig.~\ref{Fig:SHist}(b) shows a two-dimensional histogram of the single-shot NMR signals $\Delta E_{\rm{hf}}$ measured at different $T_{\rm{Evol}}$. A cross-section at short $T_{\rm{Evol}}\approx1~\mu$s is shown in Supplementary Fig.~\ref{Fig:SHist}(c), while the result for a long $T_{\rm{Evol}}\approx30$~ms is shown in Supplementary Fig.~\ref{Fig:SHist}(a). The relative weights of the ``no-flip'' and ``spin-flip'' modes reveal the probability for the electron spin to relax. We model this relaxation probability by an exponential function of $T_{\rm{Evol}}$, to derive the spin lifetime $T_{1,\rm{e}}\approx0.58$~ms for this experiment conducted at $B_{\rm{z}}=7$~T.

Supplementary Figs.~\ref{Fig:SHist}(d)--\ref{Fig:SHist}(f) show the additional results from the same two-pulse experiment, but conducted at a reduced magnetic field of $B_{\rm{z}}=2.4$~T. In this dataset, the two ``no-flip'' outcomes merge into one mode at $\Delta E_{\rm{hf}}\approx25~\mu$eV. This overlap is due to a slight nonlinearity in the dependence of the optically-probed hyperfine shift $\Delta E_{\rm{hf}}$ on the QD nuclear spin polarization. Such nonlinearity is a result of the well-known feedback effect occurring in the electron-nuclear spin system under optical pumping \cite{Urbaszek2013}. Apart from that, the results in Supplementary Figs.~\ref{Fig:SHist}(d)--\ref{Fig:SHist}(f) match qualitatively the results in Supplementary Figs.~\ref{Fig:SHist}(a)--\ref{Fig:SHist}(c), with the  gradual emergence of the ``spin-flip'' modes at an increasing $T_{\rm{Evol}}$. However, at $B_{\rm{z}}=2.4$~T the electron spin lifetime is significantly longer, found to be $T_{1,\rm{e}}\approx5.2$~ms from the exponential model fitting. The inverse dependence of $T_{1,\rm{e}}$ on the applied magnetic field is a clear indication that the electron spin relaxation is dominated by the acoustic phonons \cite{Khaetskii2001,Gillard2021}.

The readout of the electron spin via nuclear spin environment is possible in a wide range of magnetic fields, as demonstrated in Figs.~2(b) and 2(c) of the main text, where similarly high fidelities are achieved at $B_{\rm{z}}=1.6$ and $B_{\rm{z}}=5.3$~T. At high magnetic fields the readout fidelity is fundamentally limited by the shortening of the electron spin lifetime $T_{1,\rm{e}}$. In the experiments, we have verified our readout techniques for magnetic fields up to $B_{\rm{z}}=7$~T [Supplementary Figs.~\ref{Fig:SHist}(a)--\ref{Fig:SHist}(c)].

At low magnetic fields, the readout is fundamentally limited by the backaction, as discussed in more detail in the next section. Backaction becomes particularly strong if the hyperfine shift $\approx Na$ of the nuclei used for the measurement (here $N$ is the number of the polarized nuclei rotated by the RF pulse, rather than the total number of nuclei in a QD) is comparable to or larger than the electron Zeeman splitting $h\nu_{{\rm{e}},0}$. Under these conditions, the electron spin energy splitting $h\nu_{{\rm{e}}}$, which is the sum of $h\nu_{{\rm{e}},0}$ and the hyperfine shift, can become very small or even zero at some point during the RF pulse. Then, the electron-nuclear spin flip-flops become energetically allowed, disrupting the electron spin qubit state. Such backaction can be remedied by using a smaller number of nuclei in the measurement, though the drawback is the reduced NMR signal $\Delta E_{\rm{hf}}$. In our experiments, electron spin readout has been verified down to $B_{\rm{z}}=0.98$~T.  At this magnetic field the backaction is not yet the limiting factor. Instead, the limitation comes from the reduction of the optically-pumped nuclear polarization at small magnetic fields. This reduction leads to a smaller NMR signals $\Delta E_{\rm{hf}}$, and less resolved spin-up and spin-down modes in histograms, such as shown in Figs.~2(b) and 2(c) of the main text. However, such limitation is technical rather than fundamental, since PL collection efficiency in our setup can still be improved by one or two orders of magnitude, e.g. by using a solid-immersion lens (SIL) \cite{Serrels2008}. With better PL photon collection, the statistical noise in $\Delta E_{\rm{hf}}$ can be reduced. Alternatively, a more accurate measurement of $\Delta E_{\rm{hf}}$ can be achieved through resonance fluorescence \cite{Munsch2014}, rather than PL. We therefore expect that our readout method should be applicable in GaAs QDs at least down to a few hundreds of mT.

\section{Numerical simulation of the electron-nuclear spin evolution}
\label{sec:DiffNum}

We perform exact numerical simulations on a system where the central electron spin ${\bf{s}}$ is coupled to an ensemble of $N$ nuclei via the contact hyperfine interaction (Supplementary Eq.~\ref{Eq:Hhfe}):
\begin{align}
\mathcal{H}_{\rm{hf}}=\sum_{k=1}^N{a_k(\hat{s}_{\rm{x}}\hat{I}_{{\rm{x}},k}+\hat{s}_{\rm{y}}\hat{I}_{{\rm{y}},k}+\hat{s}_{\rm{z}}\hat{I}_{{\rm{z}},k})},\label{Eq:Hhfe2}
\end{align}
The Zeeman terms are:
\begin{eqnarray}
\mathcal{H}_{\rm{Z,e}} = h\nu_{{\rm{e}},0} \hat{s}_{{\rm{z}}},\label{Eq:HZe} \\ 
\mathcal{H}_{\rm{Z,N}} =-h\sum_{k=1}^N \nu_{{\rm{N}},k} \hat{I}_{{\rm{z}},k},\label{Eq:HZN2}
\end{eqnarray}
where the nuclear Zeeman term of Supplementary Eq.~\ref{Eq:HZN} is rewritten in terms of the nuclear Larmor frequencies $\nu_{{\rm{N}},k}$. The bare electron Larmor frequency is $\nu_{{\rm{e}},0}=\mu_{\rm{B}} g_{\rm{e}} B_{\rm{z}}/h$, where $\mu_{\rm{B}}$ is the Bohr magneton and $g_{\rm{e}}$ is the electron $g$-factor. For completeness, we include the nuclear-nuclear dipolar interaction (Supplementary Eq.~\ref{Eq:HDD}). The term that describes the time-dependent radiofrequency (RF) field $\nu_{1}(t)$ acting on the nuclei is given by:
\begin{eqnarray}
\mathcal{H}_{\rm{RF,N}} =-h\sum_{k} \nu_{1}(t) \hat{I}_{{\rm{x}},k},\label{Eq:HRF}
\end{eqnarray}
where the summation goes over only those nuclei that belong to the isotope that is resonant with the RF.

Direct numerical modelling is carried out for up to $N\leq 12$ nuclei with spin $I=1/2$. We simplify the problem by assuming uniform nuclear Zeeman frequencies $\nu_{{\rm{N}},k}=\nu_{\rm{N}}$ and hyperfine constants $a_k=a$. The evolution of the system is simulated through numerical propagation of the Schrödinger equation from an initial wavefunction state $\psi_{\rm{Init}}$. The computation is carried out using the software package Wolfram Mathematica 13.2. We chose $\psi_{\rm{Init}}$ as a product state of the electron and the nuclear spins, which means that the spins are not entangled initially. Moreover, the nuclear spin ensemble is initialized into a product of identical single-nucleus states. Next, we describe numerical simulations under different settings.

\subsection{Measurement contrast and backaction on the qubit}
\label{subsec:Backaction}

First, we model the QND measurement process for the case where the electron is initially in the measurement (energy) basis eigenstate. The nuclei are initially in a fully-polarized state with $I_{{\rm{z}},k}=+1/2$ for all $k$. An RF pulse with a total duration of $T_{\rm{RF}}$ is applied. In order to match the experiments, we use a cosine amplitude envelope $\nu_1(t)\propto 1 - \cos(2\pi t/T_{\rm{RF}})$, where the proportionality factor is chosen to produce a $\pi$ rotation (inversion) of the nuclei when the RF frequency is in resonance with the nuclei. The RF frequency is $\nu_{\rm{N}}-a/(2h)$, detuned from the bare NMR frequency $\nu_{\rm{N}}$. For an electron in the $s_{\rm{z}}=+1/2$ ($s_{\rm{z}}=-1/2$) state the RF pulse is resonant (detuned), resulting in a full (partial) inversion of the nuclear spins. We note certain similarities of this spin-to-spin conversion via off-resonant NMR with the dispersive (frequency-detuned) readout of a superconducting qubit coupled to a microwave cavity \cite{Blais2004} -- the role of the bosonic cavity mode is similar to that of the fermionic nuclear spin ensemble in our simulations and experiments on QDs.

Using the wavefunction $\psi_{\rm{Fin}}$ in the final state, the final polarization of each nucleus is calculated from the expectation value $I_{{\rm{z}},k,{\rm{Fin}}}=\langle \psi_{\rm{Fin}}\vert \hat{I}_{{\rm{z}},k}\vert \psi_{\rm{Fin}}\rangle$, with a total nuclear polarization defined as $\Sigma I_{{\rm{z}},{\rm{Fin}}}=\sum_{k=1}^{N} I_{{\rm{z}},k,{\rm{Fin}}}$. The measurement contrast $\Delta\Sigma I_{{\rm{z}}}$ is the difference in $\Sigma I_{{\rm{z}},{\rm{Fin}}}$ obtained under $s_{\rm{z}}=-1/2$ and $s_{\rm{z}}=+1/2$ electron states. We conduct simulations for a wide range of parameters $T_{\rm{RF}}$, $a$, $N$, $\nu_{\rm{N}}$ and $\nu_{\rm{e},0}$, under the condition $\nu_{\rm{N}}<\nu_{\rm{e},0}$. Despite these variations, we find that the normalized contrast $\Delta\Sigma I_{{\rm{z}}}/N$ depends on a single combination $(aT_{\rm{RF}}/h)^2$, as plotted by the symbols in Supplementary Fig.~\ref{Fig:Sim}(a). For short measurement times and/or weak hyperfine interaction $(aT_{\rm{RF}}/h)^2\ll 1$, the NMR resonances under $s_{\rm{z}}=\pm1/2$ electron states are resolved only partially, leading to a partial measurement contrast $\Delta\Sigma I_{{\rm{z}}}/N\ll 1$. In the opposite limit of slow measurement $(aT_{\rm{RF}}/h)^2\geq 1$ the contrast saturates at $\Delta\Sigma I_{{\rm{z}}}/N\approx 1$, which is the regime used in our experiments. The transition from partial to full contrast is well described by the following empirical expression (plotted by the solid line):
\begin{eqnarray}
\Delta\Sigma I_{{\rm{z}}}/N \approx \left(1 + (1.418\times(aT_{\rm{RF}}/h)^2)^{-2} \right)^{-1/2}\label{Eq:DeltaJz}
\end{eqnarray}

\begin{figure*}
\includegraphics[width=0.99\linewidth]{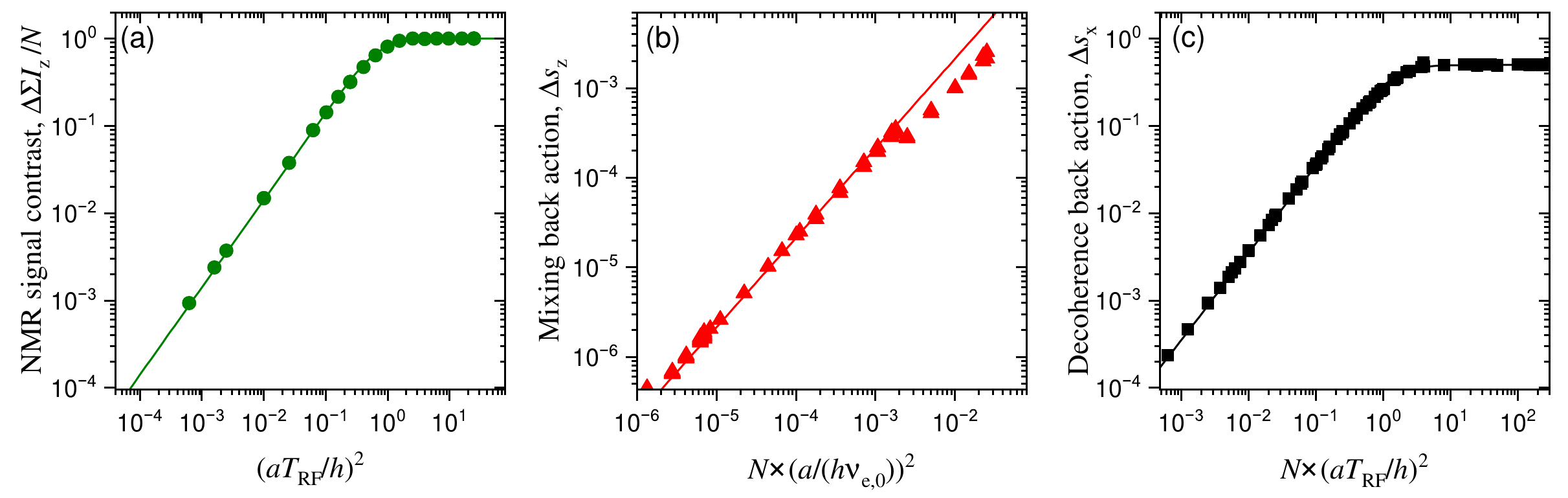}
\caption{Results of numerical simulations of the electron-nuclear spin dynamics. Each point in the plots is a result of a numerical modelling for a certain set of parameters $T_{\rm{RF}}$, $a$, $N$, $\nu_{\rm{N}}$ and $\nu_{\rm{e},0}$. These parameters are varied in a wide range. The results are plotted as a function of the parameter combination that yields a universal functional dependence. (a) QND measurement contrast defined as a difference in the final nuclear spin polarizations $\Delta\Sigma I_{{\rm{z}}}$ produced by the spin up and spin down electrons. The normalized measurement contrast is plotted as a function of $(aT_{\rm{RF}}/h)^2$. (b) Backaction in the form of mixing, defined as the RF-induced variation in the electron spin $z$ projection. The horizontal axis is $N \left( \frac{a}{h\nu_{{\rm{e}},0}} \right)^2$. (c) Backaction in the form of decoherence, defined as the RF-induced variation in the electron spin $x$ projection. The horizontal axis is $N(aT_{\rm{RF}}/h)^2$.} \label{Fig:Sim}
\end{figure*}

The contrast $\Delta\Sigma I_{{\rm{z}}}/N$ is the desired effect produced by the RF measurement pulse. We now evaluate the undesired effect of the same RF pulse, namely the disturbance of the electron spin qubit state. Following the RF pulse, the final electron spin polarization is calculated as $s_{\rm{z,Fin}}=\langle \psi_{\rm{Fin}}\vert \hat{s}_{\rm{z}}\vert \psi_{\rm{Fin}}\rangle$. Its deviation $\Delta s_{\rm{z}}$ from the initial $s_{\rm{z}}$ is a measure of backaction in the form of mixing between the measurement basis states. If the electron is in the $s_{\rm{z}}=-1/2$ state, $\Delta s_{\rm{z}}$ is found to be small. This is because the NMR is shifted out of resonance by the electron Knight field and the nuclei are not flipped. In other words, the RF pulse causes minimal evolution of the electron-nuclear spin system. For the opposite electron spin state $s_{\rm{z}}=+1/2$ the backaction $\Delta s_{\rm{z}}$ is found to be larger. In this case the nuclei are flipped by the RF pulse, and the transient hyperfine (Overhauser) field produced by the nuclei is what causes the backaction on the electron spin. Consequently, we focus on the worst-case scenario of $s_{\rm{z}}=+1/2$.  Once again, for a wide range of model parameter we find a universal functional dependence:
\begin{eqnarray}
\Delta s_{\rm{z}} \approx 0.214 \times N \left( \frac{a}{h\nu_{{\rm{e}},0}} \right)^2,\label{Eq:DeltaSz}
\end{eqnarray} 
as can be seen in Supplementary Fig.~\ref{Fig:Sim}(b). The mixing backaction is seen to be a perturbative effect, which vanishes when the electron-nuclear coupling $\propto a$ is small compared to the electron spin energy gap $h\nu_{{\rm{e}},0}$. Extrapolating the exact results obtained at $N\leq 12$ we estimate $\Delta s_{\rm{z}} $ for our experiments on GaAs QDs. The typical electron spin splitting is $h \nu_{\rm{e}}\approx50~\mu$eV, arising both from the Zeeman splitting $h\nu_{{\rm{e}},0}$ and the polarized nuclei that are not used in the electron spin readout. We note that the numerator in Supplementary Eq.~\ref{Eq:DeltaSz} can be rearranged as $Na^2=a \times (Na)$. The typical Knight shift is  $a/(2h)\approx70$~kHz. The product $Na$ is the electron hyperfine (Overhauser) splitting due to the nuclei that are used in the electron spin readout (i.e. rotated by the RF pulse), and is typically $\lesssim 15~\mu$eV. Substituting these numbers we find $\Delta s_{\rm{z}}\approx6\times 10^{-7}$. The smallness of the backaction on the measured variable $s_{\rm{z}}$ is a key defining property of the quantum non-demolition (QND) measurement \cite{Blais2004,Ralph2006}. The backaction $\Delta s_{\rm{z}}$ can be interpreted as QND infidelity $1-F_{\rm{QND}}$. The estimated $\Delta s_{\rm{z}}$ is also small compared to the overall measurement infidelity $1-F\approx 1-0.9985$, confirming that backaction is not the limiting factor, and that the sub-unity fidelity is caused by the electron-spin qubit relaxation.

Conjugate to the observable $s_{\rm{z}}$ is the electron spin coherence, which can be written in terms of the azimuthal angle \cite{Dubois2023} of the electron spin in the $xy$ plane. In order to evaluate the measurement backaction in the form of electron spin qubit decoherence we perform the same numerical simulations of the QND process. The only difference is that the electron is now initialized in a superposition of the spin up and down states $\psi_{\rm{Init}}=2^{-1/2} (\vert+1/2\rangle + \vert-1/2\rangle)$, which is the $s_{\rm{x}}=+1/2$ eigenstate. Following the RF pulse, the $s_{\rm{x,Fin}}=\langle \psi_{\rm{Fin}}\vert \hat{s}_{\rm{x}}\vert \psi_{\rm{Fin}}\rangle$ expectation is calculated to find the deviation $\Delta s_{\rm{x}}$, which characterizes the degree of the measurement-induced electron spin qubit decoherence. The universal functional dependence is found to be of the form:
\begin{eqnarray}
\Delta s_{{\rm{x}}} \approx \left((1/2)^{-2} + (0.353\times N(aT_{\rm{RF}}/h)^2)^{-2} \right)^{-1/2},\label{Eq:DeltaSx}
\end{eqnarray}
as shown in Fig.~\ref{Fig:Sim}(c). There are two distinct cases. In the limit of short measurement $N(aT_{\rm{RF}}/h)^2\ll 1$ the decoherence $\Delta s_{{\rm{x}}}$ scales quadratically with $aT_{\rm{RF}}/h$. In case of one nuclear spin ($N=1$), the decoherence $\Delta s_{{\rm{x}}}$ can be interpreted as the phase acquired by the electron spin through its interaction with the nucleus over the measurement time $T_{\rm{RF}}$. Note that $N(aT_{\rm{RF}}/h)^2\ll 1$ implies $(aT_{\rm{RF}}/h)^2\ll 1$. Taking into account Supplementary Eq.~\ref{Eq:DeltaJz}, this means that a small decoherence $\Delta s_{{\rm{x}}}$ can be achieved only at the expense of a reduced measurement readout contrast $\Delta\Sigma I_{{\rm{z}}}$ (i.e. reduced measurement fidelity). This is the case of a weak measurement, where the amount of information obtained is small and the backaction is also small for both the measured variable and the conjugate variable \cite{Ralph2006,Hatridge2013,Cujia2019,Pfender2019}. In the opposite case $N(aT_{\rm{RF}}/h)^2\geq 1$, the decoherence is nearly complete $s_{\rm{x,Fin}}\approx0$, $\Delta s_{{\rm{x}}}\approx1/2$. Moreover, we find that the expectation values of the other electron spin projections also vanish: $s_{\rm{y,Fin}}\approx s_{\rm{z,Fin}}\approx0$. This indicates that electron spin coherence is lost through the measurement-induced entanglement of the electron spin with the nuclear spins, rather than through coherent electron spin precession. Substituting the experimental parameters, we find $N(aT_{\rm{RF}}/h)^2\approx 8\times10^5\gg1$, indicating that our quantum-dot measurement of $s_{\rm{z}}$ is associated with a complete loss of the conjugate variable (i.e. complete qubit decoherence). In other words, our experiments can be described as a strong QND measurement.

\subsection{Ruling out the measurement-induced ``wavefunction collapse''}
\label{subsec:Linearity}

Experiments, such as those shown in Fig.~3(b,c) of the main text, indicate that in the vast majority of the single-shot measurements the electron is detected in either of its two energy eigenstates $s_{\rm{z}}=\pm1/2$. This observation brings up the following questions: Why is it that coherent superpositions are not observed? Does the measurement itself cause the ``collapse'' of the wavefunction, projecting the electron spin onto the $s_{\rm{z}}$ energy basis? At present, it is not possible to create coherent electron spin states in our experiments. Therefore, we use numerical modelling to study these regimes. Once again, we model the QND measurement process with nuclei initialized in a fully-polarized state with $I_{{\rm{z}},k}=+1/2$ for all $k$. The electron spin is initialized in a general superposition $\psi_{\rm{Init}}=\alpha \vert+1/2\rangle + \beta \vert-1/2\rangle$ with real $\alpha$ and $\beta$, which corresponds to an electron spin eigenstate in the $xz$ plane. The $z$-projection expectation value is initially $s_{\rm{z,Init}}=(\vert\alpha\vert^2-\vert\beta\vert^2)/2$, while the initial $x$-projection expectation is $s_{\rm{x,Init}}=\alpha\beta$. In this calculation we use $N=10$, $a/(2h)=100$~kHz, $\nu_{\rm{N}}=2$~MHz, $T_{\rm{RF}}=25~\mu$s and $\nu_{\rm{e},0}=20$ or 200~MHz. These parameter sets correspond to the case of the well-resolved Knight-shifted NMR resonances ($aT_{\rm{RF}}/h>1$) where the measurement contrast $\Delta\Sigma I_{{\rm{z}}}$ is close to its maximum value. Following the detuned RF pulse we evaluate the expectation value of the spin $z$ projection of the $k$-th nucleus: $I_{{\rm{z}},k,{\rm{Fin}}}=\langle \psi_{\rm{Fin}}\vert \hat{I}_{{\rm{z}},k}\vert \psi_{\rm{Fin}}\rangle$. With good accuracy we find that spin polarization of each nucleus replicates the initial electron spin polarization $I_{{\rm{z}},k,{\rm{Fin}}}\approx s_{\rm{z,Init}}$. This result can be understood qualitatively by noting the large difference in the nuclear and electron spin precession frequencies $\nu_{\rm{N}}\ll\nu_{\rm{e},0}$, meaning that the fast electron spin precession is averaged out and the nuclei effectively sense only the average polarization $s_{\rm{z,Init}}$ of the electron spin. 

Moreover, the electron spin $z$ polarization is essentially unchanged by the measurement RF pulse $s_{\rm{z,Fin}}\approx s_{\rm{z,Init}}$, as expected for a QND measurement. This observation allows us to rule out the possibility that the measurement RF pulse can itself ``collapse'' the electron wavefunction onto the $s_{\rm{z}}$ eigenbasis. On the other hand, regardless of the initial $s_{\rm{x,Init}}$, the final transverse electron spin components are found to be small $s_{\rm{x,Fin}}\approx s_{\rm{y,Init}}\approx0$, signifying electron spin decoherence through entanglement with the nuclear spins.


\subsection{Ruling out the nuclei as a source of einselection}
\label{subsec:Einselection}

Histograms of the experimentally measured single-shot NMR signals [Figs.~3(b) and 3(c) of the main text] reveal sharp bimodal distributions of $\Sigma I_{{\rm{z}},{\rm{Fin}}}$, indicating that the electron is found preferentially in the $s_{\rm{z}}=\pm1/2$ energy eigenstates. In order to verify this interpretation, we use the same model parameters as in Fig.~3(c), but assume a uniform distribution of the electron spin vector on a Bloch sphere (i.e. assume that the electron is in a random superposition of the $s_{\rm{z}}=\pm1/2$ states). Given the linear response of the measurement $I_{{\rm{z}},k,{\rm{Fin}}}\approx s_{\rm{z,Init}}$ established above (\ref{subsec:Linearity}), the single-shot NMR signals $\Delta E_{\rm{hf}}$ should have the same distribution as $s_{\rm{z}}$. In other words, the NMR measurement is in principle capable of detecting the electron spin superposition states. The distribution calculated under the assumption of uniformly distributed electron spin superpositions is plotted by the dashed line in Fig.~3(c) of the main text. It is nearly uniform and is incompatible with the experimental histogram, confirming that electron spin superpositions are not realized under equilibrium conditions in our experiments. Quantum mechanics gives no a priori preference to the energy eigenbasis or any other basis. Such a preferential basis, to which the system decoheres from a superposition, can arise from the interaction of the qubit with the environments. This phenomenon is known as einselection \cite{Schlosshauer2005,Zurek2018}. The bimodality of the NMR readouts confirms that such einselection takes place in the experiments. However, the results of the numerical modelling presented in \ref{subsec:Linearity} rule out RF manipulation of the nuclear spins as a mechanism of einselection. 


In order to complete this analysis, we consider the possibility of einselection induced by the slow equilibrium electron-nuclear spin dynamics, as opposed to the case of fast spin dynamics during the short (tens of $\mu$s) RF measurement pulse, as considered above. Here, we construct the initial wavefunction $\psi_{\rm{Init}}$ as a product state, where the nuclei and the electron are initially polarized in the $xy$ plane (which is orthogonal to the static magnetic field direction $z$). The nuclei are aligned along the $x$ axis, so that each nucleus is initially in a superposition $2^{-1/2} (\vert+1/2\rangle + \vert-1/2\rangle)$ of its single-particle eigenstates  $I_{{\rm{z}},k}=\pm1/2$. We consider different initial orientations of the electron spin, by taking the initial superposition of the $s_{{\rm{z}}}=\pm1/2$ states as $2^{-1/2} (\vert+1/2\rangle \pm \vert-1/2\rangle)$ (corresponding to the $s_{{\rm{x}}}=\pm1/2$ eigenstate) or $2^{-1/2} (\vert+1/2\rangle \pm i \vert-1/2\rangle)$ (corresponding to the $s_{{\rm{y}}}=\pm1/2$ eigenstate). For all these states the $z$ projection expectation value is $s_{\rm{z,Init}}=0$. There is no RF pulse in this simulation -- the electron-nuclear system is allowed to evolve freely for a few tens of milliseconds, which is much longer than all the relevant interaction timescales, and is therefore sufficient to achieve the steady state. We then calculate the final electron spin polarization $s_{\rm{z,Fin}}=\langle \psi_{\rm{Fin}}\vert \hat{s}_{\rm{z}}\vert \psi_{\rm{Fin}}\rangle$ that emerges from the electron-nuclear interaction. The value of $s_{\rm{z,Fin}}$ depends on the initial mutual orientation of the electron and nuclear spins. In case of the orthogonal orientation ($s_{{\rm{y}}}=\pm1/2$) we find $s_{\rm{z,Fin}}\approx0$. In case when the electron spin $s_{{\rm{x}}}=-1/2$ ($s_{{\rm{x}}}=+1/2$) is initially aligned (anti)parallel to the nuclei, we find negative (positive) $s_{\rm{z,Fin}}$. For a wide range of parameters we find that the magnitude follows an empirical relation:
\begin{eqnarray}
\vert s_{\rm{z,Emerg}} \vert \approx 0.231\times N \frac{a}{h\nu_{{\rm{e}},0}}.\label{Eq:EmergSz}
\end{eqnarray}
Up to a factor on the order of unity, this result has a simple interpretation as a ratio of two energies. The initial mutual electron-nuclear hyperfine energy is $\pm Na/4$ according to Supplementary Eq.~\ref{Eq:Hhfe2}, while the electron spin energy splitting is $h\nu_{{\rm{e}},0}$. We now extrapolate these results to the case of a real GaAs QD. In a thermal equilibrium any non-zero nuclear spin polarization is due to the statistical fluctuations, which scale as $\approx\sqrt{N}$. Thus we substitute $N$ with $\sqrt{N}$ and $\nu_{{\rm{e}},0}$ with $\nu_{{\rm{e}}}$ in Supplementary Eq.~\ref{Eq:EmergSz}, and use the realistic values $h \nu_{\rm{e}}\approx50~\mu$eV,  $a/(2h)\approx70$~kHz, $N\approx10^5$ to find a small final nuclear spin polarization $\vert s_{\rm{z,Emerg}} \vert \approx10^{-3}$. The emergence of a small $\vert s_{\rm{z,Emerg}} \vert\ll1/2$ has a simple interpretation in that the hyperfine energy of an equilibrium nuclear spin fluctuation is much smaller than the electron Zeeman energy, making it energetically impossible for the nuclei to ``collapse'' the electron spin superposition into its energy eigenbasis. Thus we rule out the low-energy nuclear spin environment as a source of einselection for the electron spin.

Unlike the low-energy nuclear spins, the crystalline environment of the QD electron can act as a high-energy environment responsible for einselection. Coupling between the QD electron spin and the phonons manifests in electron spin relaxation. At low temperatures ($T=4.2$~K), the electron spin relaxation is dominated by single-phonon processes \cite{Khaetskii2001,Gillard2021}. Reduction of the electron spin lifetime $T_{\rm{1,e}}$ with the increasing magnetic field $B_{\rm{z}}$ indicates that the phonon coupling is the dominant electron spin relaxation channel. By contrast, relaxation of the QD electron spin due to its cotunneling coupling with the Fermi reservoir would have resulted in $B_{\rm{z}}$-independent spin lifetimes \cite{Gillard2021}. Following Eq.~4 of Ref.~\cite{Khaetskii2001} the effective spin-phonon coupling can be written as $\propto(\hat{s}_{\rm{x}}\mathcal{E}_{\rm{y}}-\hat{s}_{\rm{y}}\mathcal{E}_{\rm{x}})$, where $\mathcal{E}_{\rm{x,y}}$ are the Cartesian components of the phonon-induced piezo-strain electric field. Such form is akin to the magnetic spin resonance Hamiltonian $\propto(\hat{s}_{\rm{x}}B_{\rm{x}}+\hat{s}_{\rm{y}}B_{\rm{y}})$. Therefore, we expect that only the resonant or nearly-resonant spectral components of $\mathcal{E}_{\rm{x,y}}$ can cause the electron spin flips. The two-pulse QND measurement experiments, shown in Supplementary Fig.~\ref{Fig:SHist} and in the main text, indicate that electron spin relaxation is well described by a telegraph random process. The phonon-induced microwave electric fields, which drive this telegraph process, must therefore occur in the form of short bursts (much shorter than the RF pulse $\lesssim10~\mu$s), separated by long (milliseconds) random intervals. Such electric field bursts can rotate the electron spin at random times and with random phases, which would explain both the spin relaxation and the einselection. Spontaneous collapses and burst-like revivals have long been investigated in Bosonic system, such as photons \cite{Eberly1980} and phonons \cite{Hizhnyakov1996,Misochko2004}, and are typically associated with high mode population numbers $\bar{n}\gtrsim 100$. For our experiments at $T=4.2$~K and $h\nu_{\rm{e}}\approx50~\mu$eV, the average phonon number is $\bar{n}\approx6.8$. The appearance of spontaneous revivals at such low excitations is somewhat unexpected and calls for further investigation.



\end{document}